\documentclass[aps,prb,twocolumn,showpacs,notitlepage,superscriptaddress]{revtex4-1}
\usepackage{graphicx,amsmath,mathrsfs}
\usepackage{amssymb}
\usepackage{amsthm}
\usepackage{bm}
\usepackage{float}
\usepackage{graphicx}
\usepackage{amsmath}
\usepackage{amstext}
\usepackage{amssymb}

\usepackage{mathrsfs}
\usepackage{amsfonts}
\usepackage{amsbsy} 
\usepackage{dcolumn}
\usepackage{bm}
\usepackage{multirow}
\usepackage{color}
\usepackage{datetime}

\usepackage{hyperref}

\usepackage{epsfig}

\usepackage{color}
\usepackage{graphicx}
\usepackage{amsmath}
\usepackage{amstext}
\usepackage{amssymb}
\usepackage{amsfonts}
\usepackage{amsbsy} 
\usepackage{dcolumn}
\usepackage{bm}
\usepackage{multirow}

\usepackage{color}
 
\usepackage[bbgreekl]{mathbbol}
\usepackage{amscd}
\def\bc{\begin{center}}
\def\ec{\end{center}}

\newcommand{\Gru}{{Gr\"uneisen}}
\newcommand{\T}{F}
\newcommand{\tf}{f}
\newcommand{\thetat}{\theta_f}
\newcommand{\VEC}[1]{{\boldsymbol{ #1}}}

\begin{document}
\title{Anharmonic interatomic force constants and thermal conductivity from \Gru{} parameters: an application to graphene}
\author{Ching Hua Lee}
\email{calvin-lee@ihpc.a-star.edu.sg}
\affiliation{Institute of High Performance Computing, 1 Fusionopolis Way, \#16-16 Connexis, Singapore 138632}
\author{Chee Kwan Gan}
\email{ganck@ihpc.a-star.edu.sg}
\affiliation{Institute of High Performance Computing, 1 Fusionopolis Way, \#16-16 Connexis, Singapore 138632}

\date{\today}
\begin{abstract}
Phonon-mediated thermal conductivity, which is of great technological
relevance, fundamentally arises due to anharmonic scattering from interatomic potentials. Despite its prevalence, accurate first-principles calculations of thermal conductivity remain challenging, 
primarily due to the high computational cost of anharmonic interatomic force constant (IFCs) calculations.
Meanwhile, the related anharmonic phenomenon of thermal expansion is much more
tractable, being computable from the \Gru{} parameters associated with phonon frequency shifts due to crystal deformations. 
In this work, we propose a novel approach for computing the largest cubic IFCs from the \Gru{} parameter data. This allows an approximate determination of the thermal conductivity via a much less expensive route. The key insight is that although the \Gru{} parameters cannot possibly contain all the information on the cubic IFCs, being derivable from spatially uniform deformations,  they can still unambiguously and accurately determine the largest and most physically relevant ones. By fitting the anisotropic \Gru{} parameter data along judiciously designed deformations, we can deduce (i.e., reverse engineer) the dominant cubic IFCs and estimate three-phonon scattering amplitudes. We illustrate our approach by explicitly computing the largest cubic IFCs and thermal conductivity of graphene, especially for its out-of-plane (flexural) modes that exhibit anomalously large anharmonic shifts and thermal conductivity contributions. 
Our calculations on graphene not only exhibits reasonable agreement with established density-functional theory results, but also presents a pedagogical opportunity for introducing an elegant analytic treatment of the \Gru{} parameters
of generic two-band models. 
Our approach can be readily extended to more complicated crystalline materials
with nontrivial anharmonic lattice effects.
\end{abstract}
\maketitle

\section{Introduction}


The harmonic approximation is ubiquitous in physics and engineering~\cite{Born56-book,Maradudin65-book}. 
Based on the Taylor expansion about an
equilibrium, it describes the leading-order oscillatory physics in
extremely diverse settings\cite{Maradudin65-book,Horton74-book}. Indeed, in much of condensed matter physics and
materials engineering, it plays a fundamental role in describing
vibrational degrees of freedom known as phonons, which underpins phenomena
like acoustic behavior, infrared and Raman spectra, and 
superconductivity\cite{Maradudin65-book}. More recently, mechanical
systems with harmonic elements have also been intensely investigated
for their extremely experimentally accessible topological
properties\cite{salerno2014dynamical,nash2015topological,wang2015topological,susstrunk2015observation,yang2015topological,zhu2015topologically,fleury2015floquet,paulose2015selective,ong2016transport,huber2016topological,liu2016topological,lu2016observation,lee2017dynamically}.

However, there are also many important phenomena based on physics
beyond the harmonic approximation\cite{Leibfried61v12,Horton74-book}. Everyday-life observations
like thermal expansion and conduction arise due to the anharmonicity of the
interatomic potential, since a purely harmonic crystal with vanishing energy
derivatives beyond the second order can neither expand nor dissipate. 
The performance of materials for heat dissipation, 
heat transport, thermal coating, and thermoelectric applications 
depends on the key quantity known as thermal conductivity\cite{Broido05v72,Snyder08v7,Ward09v80,Garg11v106,Zebarjadi12v5,Broido12v86}.
Indeed, the development of advanced technological applications like nanotube-based
electronic devices require an accurate understanding of how
phonon anharmonicity -- intrinsic or artificially induced -- affect their
thermal performance\cite{Wang07v75}.

In view of the prevalence of anharmonicity in describing these diverse
physical phenomena, various methods have been developed to calculate the
third-order derivative of energy with respect to atomic displacements,
i.e., the cubic IFCs or simply the cubic force constants (CFCs) of a crystal. We
note that techniques to obtain the second-order derivatives of energy
(or simply force constants) are well-developed, and include the supercell
method\cite{Frank95v74,Kresse95v32,Ackland97v9,Parlinski97v78,Alfe09v180,Gan06v73,Gan10v49,Liu14v16,Togo15v108,Togo15v91}
and density-functional perturbation theory (DFPT)\cite{Baroni87v58,Baroni01v73}. The
extension of the supercell methods to the third order has been
carried out in the real space\cite{Esfarjani08v77,Shiomi11v84} or in
reciprocal space using DFPT\cite{Debernardi95v75,Ward09v80,Paulatto13v87} or compressed
sensing\cite{Zhou14v113}. Other methods include extracting CFCs from molecular dynamics
simulations\cite{Hellman13v88}. Most of these methods for the computation of
CFCs and hence thermal conductivity are computationally intensive,
and it remains an open task to find more inexpensive approaches.

Meanwhile, there exists a useful measure of phonon anharmonicity known as the \Gru{}
parameter that can be easily computed from first-principles without the
explicit knowledge of the CFCs. This has been demonstrated in
Refs. \onlinecite{Pavone93v48,Mounet05v71}.
Defined as the fractional change in phonon
mode energy per fractional change in volume, it is directly related
to the thermal expansion coefficient (TEC) and the specific heat
capacity. Indeed, the determination of \Gru{} parameter has already provided
an efficient determination of thermal expansion coefficients of 
orthorhombic systems\cite{Gan15v92,Liu17v121}, a trigonal system\cite{Arnaud16v93}, and hexagonal systems\cite{Mounet05v71,Ding15v5,Gan16v94}.

In this work, we introduce an approach for computing the most
physically relevant CFCs, and hence a good approximation to the thermal conductivity, from \emph{anisotropic} \Gru{} parameters corresponding to
a set of strategically selected deformations.  Such anisotropic \Gru{} parameters have been previously employed for the calculation of thermal expansion coefficients\cite{Gan15v92,Gan16v94}. Their computation requires only knowledge of how the phonon dispersions change under spatially uniform deformations, which are obtainable through standard phonon calculations. Our approach is much less expensive than a direct evaluation of third-order derivatives of energy with respect to displacements based on supercell methods or DFPT methods~\cite{michel2015theory}, which involve meticulously keeping track of atomic displacements within the supercell.

Although the \Gru{}
parameter data so obtained clearly contains less information than those of third-order
DFPT, the key insight is that it is already sufficient for determining
the \emph{largest few} and hence most physically important CFCs and their counterparts related by symmetry. This can be systematically performed by ``reverse engineering"
the CFCs based on a general expression for the anisotropic \Gru{} parameter, to be derived in Section II, and the
\Gru{} parameters data from phonon dispersions calculated using standard density-functional theory (DFT).
We demonstrate the veracity of our approach by showing that the thermal conductivity of out-of-plane acoustic phonons for 
graphene computed from our CFCs agree fairly closely with established results.

The paper is structured as follows. We begin in Section II by developing a framework for describing crystal anharmonicity in terms of the \emph{anisotropic} \Gru{} parameters, and showing how the CFCs contained therein affect thermal expansion and conduction. Next in Section III, we detail our recipe for recovering the dominant CFCs from \Gru{} parameter data obtained via simple DFT computations on deformed crystals. In Sect IV, we provide a pedagogical illustration of how our approach can be applied to compute the CFCs and thermal conductivity of the flexural modes of graphene. 

\section{Anharmonicity in crystals}

Consider a lattice of atoms interacting via a position-dependent potential energy
\begin{eqnarray}
U=&&\frac1{2}\sum_{l,l',j,j'}\sum_{\alpha\beta}\Phi_{l j,l' j'}^{\alpha\beta}u^\alpha_{lj} u^\beta_{l'j'}\notag\\&&
+\frac1{6}\sum_{l,l'l'',j,j'j''}\sum_{\alpha\beta\gamma}\Psi_{l j,l' j',l'' j''}^{\alpha\beta\gamma}u^\alpha_{lj} u^\beta_{l'j'}u^\gamma_{l''j''}\notag\\
\end{eqnarray}
where $u^\alpha_{lj}$ is the $\alpha$th Cartesian direction
of the displacement from equilibrium of the $j$th atom in the
$l$th unit cell at $\VEC{R}_l$. $U$ comprises harmonic energy penalties from the
quadratic interatomic force constants $\Phi_{l j,l' j'}^{\alpha\beta}$
, as well as \emph{anharmonic} terms from the
CFCs $\Psi_{l j,l' j',l'' j''}^{\alpha\beta\gamma}$. Terms of quartic order can usually be neglected, except for specific materials with unusually large mean square atomic displacements\cite{tadano2015self}. In general, $\Phi_{l j,l' j'}^{\alpha\beta}$ and $\Psi_{l j,l' j',l'' j''}^{\alpha\beta\gamma}$ are not directly dependent on each other, other geometric relations exist between them\cite{Leibfried61v12}. Due
to translational invariance, we can define a dynamical matrix
\begin{equation}
D^{\alpha\beta}_{jj'}(\VEC{q})=\frac1{\sqrt{M_j M_{j'}}}\sum_{l'}\frac{\partial^2 U}{\partial u^\alpha_{0j}\partial u^\beta_{l'j'}}e^{i\VEC{q} \cdot \VEC R_{l'}},
\label{dynamical}
\end{equation}
at a wave vector $\VEC{q}$, whose squared eigenvalues $\omega_{\VEC{q}s}^2$ describe the phonon 
dispersion of the $s$th phonon branch, where $M_j$ is the mass of the $j$th atom. 

In the absence of anharmonicity, $\Phi^{\alpha\beta}_{l j,l' j'}$ is a constant that does not depend on the displacements $u^\alpha_{lj}$. In this case, then, the phonon dispersions are independent of any changes in the size or shape of the crystal\footnote{In fact, the assumption of \emph{exclusively} quadratic force constants leads to the paradoxical situation of the phonon bands remaining unchanged under arbitrarily large lattice distortion.}, precluding thermal expansion due to energetic considerations. At the many-body level, the lack of higher order terms also precludes any scattering processes that lead to thermal resistivity.

As such, a realistic phonon potential must contain anharmonic terms. 
Below, we derive an expression for the anisotropic \Gru{} parameter based on first-order perturbation theory. 
For this, we need to investigate how the phonon frequency $\omega_{\VEC{q}s}$ 
of wave vector $\VEC{q}$ and mode index $s$ changes when the crystal is deformed according to the deformation matrix $E$ given by
\begin{equation}
E = 
\left( \begin{array}{ccc} e_1 & e_6/2 & e_5/2  \\ e_6/2 & e_2 & e_4/2 \\ e_5/2 & e_4/2 & e_3 \end{array} \right)
\end{equation}
with the six parameters $e_i$, $i=1,2, \cdots, 6$ appearing as in the Voigt notation\cite{Nye85-book}.

Without loss of generality, we parametrize a small deformation as $e_i = \eta \tf_i$, $i = 1, \cdots, 6$, where $\eta$ is a small parameter and $\tf_i$ are constants normalized according to $\sum_i \tf_i^2=1$. 
Now $E = \eta \T$ where
\begin{equation}
\T =
\left( \begin{array}{ccc} \tf_1 & \tf_6/2 & \tf_5/2  \\ \tf_6/2 & \tf_2 & \tf_4/2 \\ \tf_5/2 & \tf_4/2 & \tf_3 \end{array} \right)
\end{equation}
An infinitesimal strain $\eta \T$ may result in a change in volume as atoms at 
equilibrium positions $\VEC r$ are shifted to $(\mathit{I}+\eta \T)\VEC r=\VEC r + \eta \T\VEC r$ where $\mathit{I}$ is the 
$3\times 3$ identity matrix.
Since this displacement
is much smaller than the lattice constant, the effect of anharmonicity and hence the \Gru{} parameter can be perturbatively treated
around the 
equilibrium volume (i.e., $\eta = 0$) as follows:
\begin{widetext}
\begin{eqnarray}
\gamma_{\VEC{q} s} (\T) &=& -\frac{1}{\omega_{\VEC{q} s}}\frac{\partial \omega_{\VEC{q} s}}{\partial \eta} 
= -\frac1{2\omega^2_{\VEC{q} s}}\frac{\partial (\omega^2_{\VEC{q} s})}{\partial \eta}
=- \frac1{2\omega^2_{\VEC{q} s}}\frac{\partial }{\partial \eta}\sum_{\alpha j,\beta j'}[\epsilon_{\VEC{q}s}^{\alpha j}]^* D^{\alpha\beta}_{j j'}(\VEC{q})
\epsilon_{\VEC{q}s}^{\beta j'}\notag\\
&=&- \frac1{2\omega^2_{\VEC{q} s}}\frac{\partial }{\partial \eta}
\sum_{\alpha j,\beta j'}
[ 
\epsilon_{\VEC{q}s}^{\alpha j}
]^*\left[\sum_{l'}   
\frac{1}{\sqrt{M_j M_{j'}}} 
e^{i\VEC{q} \cdot  \VEC R_{l'}} \frac{\partial^2 U}{\partial u_{0j}^\alpha\partial u_{l'j'}^\beta}\right]
\epsilon_{\VEC{q}s}^{\beta j'}
\notag\\
&=&- \frac1{2\omega^2_{\VEC{q} s}}\frac{\partial }{\partial \eta}
\sum_{\alpha j, \beta j'}
[
\epsilon_{\VEC{q}s}^{\alpha j}
]^*\left[\sum_{l'} 
\frac{1}{\sqrt{M_j M_{j'}}}
e^{i\VEC{q} \cdot  \VEC R_{l'}}\sum_{l''j''\gamma}
\left(\Psi^{\alpha\beta\gamma}_{0j,l'j',l''j''}u_{l''j''}^\gamma\right)\right]
\epsilon_{\VEC{q}s}^{\beta j'}
\notag\\
&=& -\frac1{2\omega^2_{\VEC{q} s}}
\sum_{\alpha j, \beta j'}
[
\epsilon_{\VEC{q}s}^{\alpha j}
]^*\left[\sum_{l'} 
\frac{1}{\sqrt{M_j M_{j'}}}
e^{i\VEC{q} \cdot  \VEC R_{l'}}\sum_{l''j''\gamma}
\left(\Psi^{\alpha\beta\gamma}_{0j,l'j',l''j''} \frac{\partial u_{l''j''}^\gamma}{\partial \eta} \right)\right]
\epsilon_{\VEC{q}s}^{\beta j'}
\notag\\
&=& -\lim_{\Delta \eta \rightarrow 0}\frac1{2\omega^2_{\VEC{q} s}}\frac1{\Delta \eta} \sum_{\alpha j, \beta j'} 
[ \epsilon_{\VEC{q}s}^{\alpha j} ]^*\left[\sum_{l'} \frac{1}{\sqrt{M_j M_{j'}}} 
e^{i\VEC{q} \cdot  \VEC R_{l'}}\sum_{l''j''\gamma} 
\left(\Psi^{\alpha\beta\gamma}_{0j,l'j'',l''j''}  [u_{l''j''}^\gamma|_{(\mathit{I}+  \Delta \eta \T)\VEC r}-u_{l''j''}^\gamma|_{\VEC r}   ]\right)\right] \epsilon_{\VEC{q}s}^{\beta j'} \notag\\
&=&\frac1{2\omega^2_{\VEC{q} s}}
\sum_{\alpha j, \beta j'}
[
\epsilon_{\VEC{q}s}^{\alpha j}
]^*\left[\sum_{l'} 
\frac{1}{\sqrt{M_j M_{j'}}}
e^{i\VEC{q} \cdot  \VEC R_{l'}}\sum_{l''j''\gamma\delta}\left(\Psi^{\alpha\beta\gamma}_{0j,l' j',l''j''}\T^{\gamma \delta}r_{l''j''}^\delta\right)\right]
\epsilon_{\VEC{q}s}^{\beta j'}
\label{grun0}
\end{eqnarray}
\end{widetext}
Here $\epsilon_{\VEC{q}s}^{\alpha j}$ is the polarization vector component of the $j$th atom of 
mode $\omega_{\VEC{q}s}$ in the $\alpha$th direction.
$\VEC r_{lj}=(r_{lj}^x,r_{lj}^y, r_{lj}^z)$ denotes the position of the $j$th atom within each unit cell.
We have used the relation $\sum_{\alpha j, \beta j'} [\epsilon_{\VEC{q}s}^{\alpha j}]^* D^{\alpha\beta}_{j j'}(\VEC{q})\epsilon_{\VEC{q}s}^{\beta j'}=\omega^2_{\VEC{q} s}$ which holds because the polarization vectors are defined as the eigenvectors of the (deformation dependent) dynamical matrix. 
In line three, we have discarded the quadratic contributions because they are proportional to the quadratic force constants, which are by definition constants unaffected by a deformation strain. Going to line 4, we have discarded the combined contributions from the derivatives of $[\epsilon_{\VEC{q}s}^{\alpha j}]^*$ and $\epsilon_{\VEC{q}s}^{\beta j'}$ because they disappear to first order due to the normalization of $\epsilon_{\VEC{q}s}^{\beta j'}$. 
To obtain the final line, we observe that changes in $u$, the displacement from equilibrium, are always compensated by changes in the equilibrium position $\VEC{r}$. 

Our definition of the \Gru{} parameter
depends unambiguously on the choice of $\T$ matrix. It is related to the conventional definition of the \Gru{} parameter, $\gamma'_{\VEC{q}s} $ as follows. The latter is based on the fractional change in volume 
associated with a specific deformation. Since the fractional change in volume is 
$\frac{\Delta V}{V} =  ({\rm Tr}\;\T)\eta$ 
(to linear order in $\eta$),  
we find that $
\gamma'_{\VEC{q}s} = \frac{\gamma_{\VEC{q}s}}{{\rm Tr}\;\T}$.
For a cubic crystal, we may choose $\T = \mathit{I}/3$, and the expression for the \Gru{} 
parameter is consistent with what is reported in the literature\cite{Shiomi11v84}. 

Note that due to translation invariance, Eq. \ref{grun0} should hold under arbitrary shifts of the origin $\VEC r_{l''}\rightarrow \VEC r_{l''}+\VEC r_0$, for any 
$\T$. Hence (restoring the $j$'s) $\sum_{l''}\Psi^{\alpha\beta\gamma}_{lj,l'j',l''j''}=0$ for any $l,l',j,j',j''$. Physically, this can be interpreted as Newton's third law for the CFCs: there must be no net force on an atom from all its periodic images.

The deformation matrix $\T$  can be expressed more geometrically in terms of its (weighted) 
principal axes $\VEC{\tf}_1, \cdots, \VEC{\tf}_{\Lambda}$, where $\Lambda $ is the 
number of nonzero eigenvalues. Each $\VEC{\tf}_i$ is given by $\VEC{\tf}_i=\sqrt{|\lambda_i|}\VEC{v}_i$, where $\lambda_i$ and $\VEC{v}_i$ are the $i$th eigenvalue and eigenvector of $\T$. We write
\begin{equation}
\T=\sum_{i=1}^\Lambda \xi_i \VEC \tf_i \VEC \tf_i^T,
\end{equation}
where $\xi_i=\text{sgn}(\lambda_i) = \frac{\lambda_i}{|\lambda_i|}$ depending on whether the deformation consists of tension, compression or shear; for brevity of notation, we will henceforth assume that $\xi_i=1$ throughout. 
Deformations with uniaxial, biaxial or triaxial strain correspond to $\Lambda=1$, $2$, or $3$, respectively. 

By writing the quantity in the curved parentheses of Eq. \ref{grun0} in the 
shorthand form $\sum_{\gamma\delta}[\Psi^{\alpha\beta}]^\gamma \T^{\gamma\delta}\VEC r^\delta
=\sum_i(\Psi^{\alpha\beta}\cdot \VEC{\tf}_i)( \VEC{r} \cdot \VEC{\tf}_i)$, we see 
that $\gamma_{\VEC{q} s}(\T)$ is a weighted projection of $\gamma_{\VEC{q} s}$ 
onto the $\{ \VEC{\tf}_1,\cdots,\VEC{\tf}_{\Lambda} \}$ parallelepiped. 
The decomposition into an expression of the form $\sum_i(\Psi^{\alpha\beta}\cdot \VEC{\tf}_i)( \VEC{r}\cdot \VEC{\tf}_i)$ 
suggests that the components of $\Psi^{\alpha\beta\gamma}$ are most effectively isolated by $\VEC{\tf}_i$ orthogonal to directions where the CFCs act.

\subsection{Physical manifestations of anharmonicity}
\subsubsection{Thermal expansion}

The simplest effect of anharmonicity is the ability for a crystal to
expand or shrink with increasing temperature. Existing at the level of
non-interacting phonons, it can be fully captured by the shifts of the
(single-particle) phonon branch energies due to $j$th type crystal deformation,\cite{Gan15v92,Arnaud16v93,Gan16v94} measured by
the \Gru{} parameters $\gamma_{j,\VEC{q} s}$. The linear TEC components $\alpha_i$ of a crystal with a
volume $V$ at temperature $T$ is given by\cite{Schelling03v68,Gan15v92}
\begin{equation}
\alpha_i(T) = \frac1{\Omega}C^{-1}_{ij}I_j,
\label{TEC2}
\end{equation}
\begin{equation}
I_j=\frac{\Omega}{V}\sum_{\VEC{q} s} \gamma_{j,\VEC{q} s} c_{\VEC{q} s},
\end{equation}
where $C$ is the elastic constant matrix
and $ c_{\VEC{q} s}$ is the phonon mode heat capacity, and $\Omega$ 
the equilibrium unit cell volume.  Eq. \ref{TEC2} had been used in the calculation of thermal expansion coefficients\cite{Gan15v92,Arnaud16v93,Gan16v94}, and is derived from first principles in Appendix \ref{sec:TECgru}.

\subsubsection{Thermal conductivity from acoustic phonons}

More interestingly, lattice anharmonicity also lead to thermal resistance due to phonon scattering processes. In the single-mode relaxation time approximation (SMRT), the thermal conductivity is proportional to the sum of the mode heat capacities $c_{\VEC{q}s}=\hbar\omega_{\VEC{q}s}\frac{\partial n_{\VEC{q}s}}{\partial T}$ multiplied by their squared group velocities and scattering times, i.e., 
\begin{eqnarray}
\kappa &=& \frac1{3V}\sum_{\VEC{q} s}c_{\VEC{q}s}\left(\frac{\partial \omega_{\VEC{q} s}}{\partial \VEC{q}}\right)^2\tau_{\VEC{q} s}\notag\\
&= &\frac{k_B}{3V}\sum_{\VEC{q} s}  \left(\frac{x_{\VEC{q}s}}{\sinh x_{\VEC{q}s}}\frac{\partial \omega_{\VEC{q} s}}{\partial \VEC{q}}\right)^2 \tau_{\VEC{q} s}
\label{tau1}
\end{eqnarray}
where $x_{\VEC{q}s}=\frac{\hbar \omega_{\VEC{q}s}}{2k_B T}$ and $n_{\VEC{q} s}, \tau_{\VEC{q} s}$ are respectively the Bose-Einstein occupation function and characteristic scattering time of a phonon of mode $(\VEC{q},s)$. The mode heat capacity can alternatively be expressed as $\frac{\hbar^2\omega_{\VEC{q}s}^2}{k_BT^2}n_{\VEC{q}s}(n_{\VEC{q}s}+1)$ which is proportional to the transition probability in the linearized Boltzmann equation\cite{Srivastava90-book}. 

The SMRT approximation gives the scattering time $\tau_{\VEC{q}s}$ for each mode $(\VEC{q},s)$ assuming that other modes are in equilibrium, and is generally valid at room temperature\cite{Paulatto13v87}. In the case of cubic anharmonicity, the scattering is dominated by three-phonon processes where a phonon $(\VEC{q},s)$ either decomposes into two phonons $(\VEC{q}',s')$ and $(\VEC{q}'',s'')$, or absorbs another phonon $(\VEC{q}',s')$ to form the phonon $(\VEC{q}'',s'')$ (or vice versa). The amplitudes of these processes are proportional to the squared magnitude of the 3-phonon scattering matrix elements $M_{\VEC{q}s,\VEC{q}'s',\VEC{q}''s''}$ via Fermi's golden rule, which gives the phonon broadening width\cite{Shiomi11v84,Paulatto13v87} :
\begin{eqnarray}
\tau_{\VEC{q} s}^{-1}&=& \frac{\pi}{8\hbar^2N}\sum_{\VEC{q}'s',\VEC{q}''s''}|M_{\VEC{q}s,\VEC{q}'s',\VEC{q}''s''}|^2\notag\\
&&\times[(n_{\VEC{q}'s'}+n_{\VEC{q}''s''}+1)\delta_{\omega_{\VEC{q}'s'}+\omega_{\VEC{q}''s''}-\omega_{\VEC{q}s}}\notag\\
&&+2(n_{\VEC{q}'s'}-n_{\VEC{q}''s''})\delta_{\omega_{\VEC{q}'s'}+\omega_{\VEC{q}s}-\omega_{\VEC{q}''s''}}]
\label{tau2}
\end{eqnarray} 
where $N$ is the number of modes in the Brillouin zone, and
\begin{eqnarray}
M_{\VEC{q}s,\VEC{q}'s',\VEC{q}''s''}&=&
\sum_{\alpha\beta\gamma}\sum_{j,l'j',l''j''}\Psi^{\alpha\beta\gamma}_{0 j,l' j',l'' j''}\epsilon^{\alpha j}_{\VEC{q}s}\epsilon^{\beta j'}_{\VEC{q}'s'}\epsilon^{\gamma j''}_{\VEC{q}''s''}\notag\\
&&\times \frac{e^{i(\VEC{q}'\cdot \VEC R_{l'}+\VEC{q}''\cdot \VEC R_{l''})} \delta_{\VEC{q} +\VEC{q}'+\VEC{q}''+\VEC G} }{\sqrt{M_j M_{j'}M_{j''}\omega_{\VEC{q}s}\omega_{\VEC{q}'s'}\omega_{\VEC{q}''s''}}}
\label{tau3}
\end{eqnarray}
Due to translation invariance, momentum is conserved up to a reciprocal lattice vector $\VEC G$. In Eq. \ref{tau2}, the delta functions enforce two possible types of channels of mode scattering, whose amplitudes $M_{\VEC{q}s,\VEC{q}'s',\VEC{q}''s''}$ are dictated by the projection of the CFCs onto branches $s,s'$ and $s''$ by the polarization vectors. 

The conductivity $\kappa$ will be large as long as $\tau_{\VEC{q}s}$ is large for at least one mode. In other words, $\kappa$ is dominated by the mode for which the phonon broadening $\tau^{-1}_{\VEC{q}s}$ due to scattering is the \emph{smallest}, akin to a parallel resistors model. A small $\tau^{-1}_{\VEC{q}s}$ despite large \Gru{} parameters and hence CFCs is often the result of a limited phase space for scattering\cite{lindsay2010flexural,Paulatto13v87}, as is the case of the flexural mode of graphene that we shall discuss in detail in Sect. \ref{flexuralcond}. 

\section{Reverse engineering of CFCs from \Gru{} parameters}
\label{sec:reverse}

As discussed above, the ubiquitous phenomenon of thermal resistance due to phonon mediated scattering is a higher order process whose computation requires knowledge of the individual CFCs (Eq. \ref{tau1} to \ref{tau3}). Unfortunately, existing methods to compute them are generally very expensive, as reflected by the paucity of good numerical data available.

As such, we introduce a much less numerically expensive approach for obtaining these CFCs from \Gru{} data. 
It is based on the following observations: (1) the anisotropic \Gru{} parameters 
$\gamma_{\VEC{q}s}(\T)=\gamma_{\VEC{q}s}(\VEC{\tf}_1,...,\VEC{\tf}_\Lambda )$ depend linearly on the 
CFCs $\Psi^{\alpha\beta\gamma}_{0j,l'j',l''j''}$ (Eq. \ref{grun0}), whose linear matrix equation can thus be simply inverted to yield the CFCs and, (2) this inversion is well-defined as long as not too many CFCs are included as nonzero unknowns to be determined. As seen in Fig. \ref{grapheneIFCplot}, there is typically only a handful of dominant CFCs.

The idea is to rewrite the \Gru{} parameter formula (Eq. \ref{grun0}) as a matrix equation, and invert it to obtain the CFCs from \Gru{} parameter data. In matrix form, Eq. \ref{grun0} is given by
\begin{equation}
\VEC \gamma=A\VEC \Psi,
\label{rev1}
\end{equation}
where $\VEC{\gamma}$ is a vector containing the anisotropic \Gru{} parameters $\gamma_{\VEC{q}s}(\VEC{\tf}_1,\dots,\VEC{\tf}_\Lambda )$  corresponding to linear combinations of uniaxial deformations taken in directions $\VEC{f}_1,..., \VEC{f}_\Lambda$. The components of $\VEC \gamma$ are indexed by the momentum and branch indices $\VEC{q},s$ as well as the $\Lambda (\Lambda +1)/2$ independent parameters in $\T=\sum_i^\Lambda  \xi_i \VEC{\tf}_i \VEC{\tf}_i^T$. The vector $\VEC{\Psi}$ 
analogously represent the CFC coefficients $\Psi^{\alpha\beta\gamma}_{0j,l'j',l''j''}$.
Notice that the matrix $A$, which contains the coefficients of $\Psi^{\alpha\beta\gamma}_{0j,l'j',l''j''}$ in Eq. \ref{grun0}, is in general not a square matrix. However, Eq. \ref{rev1} can still be inverted by multiplying both sides by $A^\dagger$, such that
\begin{equation}
\VEC\Psi=Q^{-1}(A^\dagger \VEC\gamma)
\label{rev2}
\end{equation}
where
\begin{equation}
A^\dagger\VEC\gamma=\sum_i 
\frac{f^\gamma_i(\VEC{f}_i\cdot \VEC r_{j''l''})}{2\sqrt{M_j M_{j'}}}
\sum_{\VEC{q} s}\frac{[\epsilon^{\beta j'}_{\VEC{q}s}]^*\epsilon^{\alpha j}_{\VEC{q}s}}{\omega^2_{\VEC{q}s}}
e^{-i\VEC{q}\cdot \VEC R_{l'}}\gamma_{\VEC{q}s}(\VEC{\tf}_i )
\end{equation}
and $Q=A^\dagger A$ is a Hermitian matrix with real non-negative eigenvalues. Explicitly, its elements are 
\begin{eqnarray}
&&Q_{\alpha_1\beta_1\gamma_1,j_1j_1'j_1'',l_1'l_1''}^{\alpha_2\beta_2\gamma_2,j_2j_2'j_2'',l_2'l_2''}
=\frac1{4}\sum_i\frac{\tf_i^{\gamma_1} \tf_i^{\gamma_2}(\VEC{\tf}_i\cdot \VEC r_{j_1''l_1''})
( \VEC{\tf}_i\cdot \VEC r_{j_2''l_2''})}{\sqrt{M_{j_1}M_{j_1'}M_{j_2}M_{j_2'}}}\notag\\
&&\times\sum_{\VEC{q}s}e^{i\VEC{q}\cdot(\VEC R_{l_1'}-\VEC R_{l_2'})}\frac{[\epsilon^{\alpha_1,j_1}_{\VEC{q}s}\epsilon^{\beta_2,j_2'}_{\VEC{q}s}]^*\epsilon^{\alpha_2,j_2}_{\VEC{q}s}\epsilon^{\beta_1,j_1'}_{\VEC{q}s}}{\omega^4_{\VEC{q} s}},
\label{Q}
\end{eqnarray}
the last line of which is a Fourier transform of the product of the four polarization vectors divided $\omega^4_{\VEC{q}s}$. Away from band degeneracies, $Q$ is thus dominated by elements with small $ l_1'-l_2'$, i.e., those close to the diagonal in the $l'$ subspace (and also the $l''$ subspace which is related by a relabeling). 

To systematically solve for the CFCs, one must ensure that $Q$ is invertible. $Q$ depends on (1) the directions of the deformations $\VEC\tf$, and (2) the choice of CFCs to be included, which enters as the atom displacement vectors  $\VEC r_{j_i''l_i''}$. As is explicit in Eq. \ref{Q}, these two types of quantities enter $Q$ via expressions $\VEC{\tf}_i\cdot \VEC r_{j_1''l_1''}$ that project $\VEC r_{j_1''l_1''}$ onto $\VEC{\tf}_i$. For $Q$ to be invertible, these projectors must be linearly independent, i.e. it must be possible to uniquely isolate the CFCs given a set of chosen\footnote{It is the combination $\sum_\gamma\Psi^{\alpha\beta\gamma}\tf_i^\gamma$, rather than the CFCs $\Psi^{\alpha\beta\gamma}$ themselves that enter the \Gru{} parameter.} $\VEC{\tf}_i$. With $\Gamma(\Gamma+1)/2$ degrees of freedom in choosing the $\VEC{\tf}_i$s, $Q$ will always be invertible when a sufficiently small set of CFCs and their symmetry-related copies are included; beyond that, the solution for the CFC coefficients will become less accurate as $Q$ becomes more singular. For our purpose, the most practical route to determining the CFCs accurately is to start from the few most local types CFCs, i.e. those with the smallest $|l|$ and $|l'|$, and then successively including the next nearest CFCs while making sure that $Q$ is not degenerate. While we a priori do not know if the most local CFCs are indeed the dominant ones, in most cases, the true magnitudes of the CFCs do decrease exponentially with distance\cite{he2001exponential,lee2016band} (See also Fig. \ref{grapheneIFCplot}). For the best convergence, it is advisable to choose the $\VEC{\tf}_i$s such that each of them is orthogonal to all but one of the position vectors $ \VEC r_{j_1''l_1''}$ of the atoms in the unit cell, since that will make the projectors within $Q$ as linearly independent as possible.


Thanks to the intrinsically local nature of atomic orbitals, the above-mentioned approach should be able to capture most of the physics despite the truncation of most three-body terms beyond the next nearest neighbor. As evident in our case study of graphene in the next section, the inclusion of five to seven types of CFCs and their various symmetry permutations is sufficient for reproducing its thermal expansion (\Gru{} parameter) and conductivity properties fairly accurately. 

To summarize, our approach for obtaining the CFCs and hence thermal conductivity involves the following steps:
\begin{enumerate}
\item Based on the crystal structure, decide on up to three independent deformation directions $\VEC{\tf}_i$ for computing the anisotropic \Gru{} parameters. Often, best results can be obtained with $\VEC{\tf}_i$s forming a subset of the dual basis to the lattice vectors within the unit cell. 
\item Choose the largest set of the most local (translation-invariant) CFCs to include in $Q$ (Eq. \ref{Q}) such that $Q$ is invertible. 
\item Compute the anisotropic \Gru{} parameters $\gamma_{\VEC{q}s}(\T)=\gamma_{\VEC{q}s}(\VEC{\tf}_1,...,\VEC{\tf}_\Lambda)$ by taking the difference in the DFT-computed phonon frequencies before and after the deformations defined by $\T=\sum_i^\Lambda  \xi_i \VEC{\tf}_i\VEC{\tf}_i^T$.
\item Obtain the CFCs via the inversion equation Eq. \ref{rev2}. 
\item If so desired, compute the thermal conductivity with these results for the CFCs via Eqs. \ref{tau1} to \ref{tau3}.
\end{enumerate}

\section{Application: Anharmonic potentials and thermal conductivity in graphene }

We devote the rest of this paper to a detailed study of phonon anharmonicity in graphene, particularly of its flexural (out-of-plane) modes. The 
high thermal conductivity of graphene has been the focus of various 
studies\cite{bonini2007phonon,Balandin08v8,Tan11v11,kong2009first}, and we shall illustrate how our approach can 
provide a reasonable estimate of its conductivity with little computational effort via DFT. Key to this high 
conductivity is the relatively restricted phase space for momentum-conserving scattering out of the flexural channel.

Graphene also makes for an excellent pedagogical example because it
can be accurately modeled as an intuitively understood generalized
spring model, also known as a minimal force-constant model.\cite{saito1998,Zimmermann08v78}
Represented this way,
its flexural degrees of freedom (DOFs) furthermore decouple from the
in-plane modes to form a simple two-band system amenable to analytic
study. As such, contributions to its conductivity from the individual
CFCs can be analytically tracked and physically interpreted in a
level of detail beyond usual thermal conductivity studies. 

\subsection{Tight-binding model set-up}

The phonon dispersions of graphene can be accurately reproduced by
a minimal force-constant tight-binding model expounded by Saito {\it et
al.}\cite{saito1998} and Zimmermann {\it et al.}\cite{Zimmermann08v78} In
such a model, graphene is represented by a honeycomb lattice of
``generalized springs" or ``elastic frames" whose restoring force $\VEC{F}$  acts
not just along the (radial) direction of stretching, but also in the
two other (perpendicular) shear directions.

\begin{figure}[H]
\includegraphics[scale=.32]{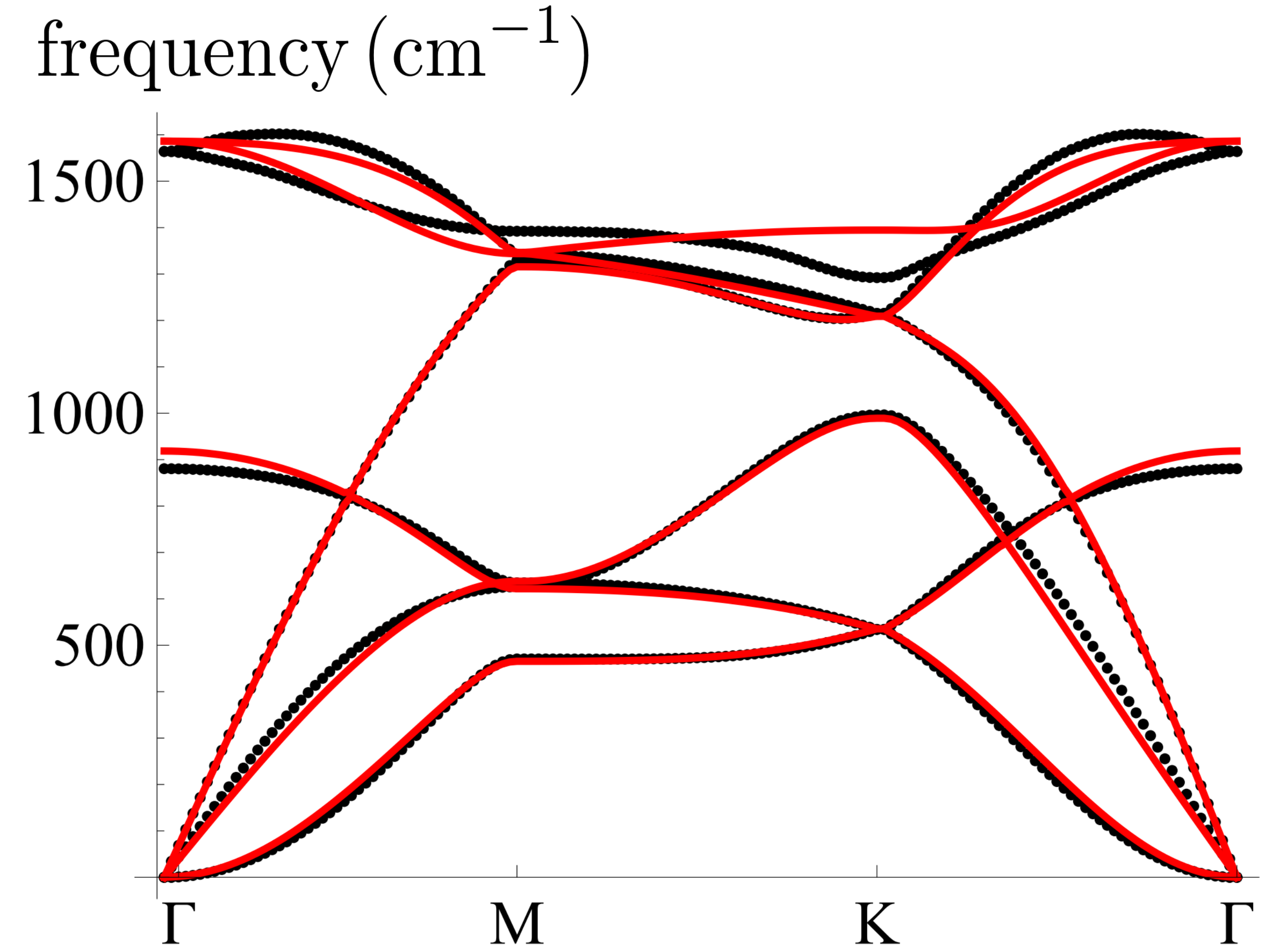}
\caption{(Color online) Comparison of the phonon dispersion of graphene computed from Eq. \ref{Force2} using tight-binding parameters from Ref. \onlinecite{Zimmermann08v78} (red), against our DFT results (black). There is excellent numerical agreement, especially for the acoustic (low frequency) modes relevant to room-temperature behavior. Near the $\Gamma$ point, the flexural modes exhibit quadratic dispersion. } 
\label{bandfitting}
\end{figure}

Consider a spring between two atoms with relative displacement $\VEC n+\Delta \VEC u$, where $\VEC n$ is their equilibrium relative displacement and $\Delta \VEC u$ the spring extension. For small extensions, the restoring force is (in the basis of $\hat x,\hat y$ and $\hat z$ displacements) given by 
\begin{eqnarray}
\VEC F
&=& - K^\parallel (\Delta \VEC u\cdot \hat{\VEC{n}})\hat{\VEC{n}} - \sum_{i=1,2} K^{\perp}_{i} (\Delta \VEC u\cdot \hat{\VEC{n}}_{\perp,i})\hat{\VEC{n}}_{\perp,i}
\label{Force}
\end{eqnarray}
where $\hat{\VEC{n}}_{\perp,i}$ are two orthogonal unit vectors perpendicular to $\VEC n$. The $K^\parallel$ term provides the restoring force in the direction along the spring, and is typically the largest. The $K^\perp_i$ terms transmits shear forces perpendicular to the spring, and exist due to the rigidity of the spring. For the purpose of studying a free-standing graphene sheet, we shall set $\hat{\VEC{n}}_{\perp,1}=\hat{\VEC{n}}_{\perp,{\rm in}}$ and $\hat{\VEC{n}}_{\perp,2}=\hat{\VEC{n}}_{\perp,{\rm out}}$ to be in-plane and out-of-plane respectively.

Recast in matrix form $\VEC F=-K\Delta \VEC u$, Eq. \ref{Force} defines a stiffness tensor $K$ given by
\begin{equation}
K= K^\parallel \hat{\VEC{n}} \hat{\VEC{n}}^T + K^\perp_{\rm in} \hat{\VEC{n}}_{\perp,{\rm in}}\hat{\VEC{n}}_{\perp,{\rm in}}^T+K^\perp_{\rm out} \hat z\hat z^T
\label{Force2}
\end{equation}
with the $3$ real parameters $K^\parallel,K^\perp_{\rm in}$ and $K^\perp_{\rm out}$ being its normal mode eigenvalues. It has three 
fewer DOFs than a generic (symmetric) stiffness tensor, which has six real DOFs, because the orientation of $\hat{\VEC{n}}$ fixes two of the Euler angles, while the stipulation of $\hat{\VEC{n}}_{\perp,2}$ being out-of-plane fixes the third one. 

The tight-binding model for graphene is constructed with the quadratic IFCs  given by $K$ in Eq. \ref{Force2}, for up to fourth
nearest neighbors. The parameters $K^\parallel,K^\perp_{\rm in}$ and $K^\perp_{\rm out}$ for each type of neighbor is taken from Ref. \onlinecite{Zimmermann08v78}, yielding an excellent fit with our DFT data, especially for the lower bands as shown in Fig. \ref{bandfitting}. 
The DFT computation is performed using the Quantum Espresso package\cite{Giannozzi09v21} with 
the same parameters as used in
Ref. \onlinecite{Mounet05v71}.
Due to the rapid spatial decay of the interatomic orbitals, there is no need to include any longer-range hoppings in Eq. \ref{Force2}. Physically, the good numerical agreement was possible because the $\pi$ orbitals between the carbon atoms are rather symmetric about their axes, and hence possess normal modes almost parallel 
or perpendicular to the direction of interatomic separation.

Most strikingly, the low-energy (acoustic) phonon behavior of graphene is governed by a pair of so-called flexural (out-of-plane) modes possessing quadratic dispersions around the zero-momentum ($\Gamma$) point. Physically, this is because energy penalties for the out-of-plane modes are due to curvature (bending) of the displacement field, rather than compressive strain\footnote{In terms of the tight-binding parameters, the lack of compressive contributions is equivalent to the out-of-plane spring constants satisfying the sum rule (written here for the honeycomb lattice) $t_{1}+6t_{2}+4t_{3}+14t_{4}+...=0$, where $t_i$ refers to $K^\perp_{\rm out}$ for the $i$th nearest neighbor.}. Mathematically, it scales like $(\nabla^2 u)^2\sim k^4 u \sim \omega^2 u$, which implies $\omega\sim k^2$. This quadratic dispersion implies a large degeneracy at low energy, which we shall see profoundly increases the size of the \Gru{} parameter as well as thermal conductivity.

An important consequence of Eq. \ref{Force} or \ref{Force2} is that, for a planar system, the flexural mode never mixes with the in-plane modes. In the case of graphene, the eigenmodes hence live in the direct sum of two decoupled subspaces: the 4-dimensional subspace of in-plane modes and the 2-dimensional (2-band) subspace of flexural modes, spanned by the two sublattices of the honeycomb lattice. The simplicity of this effective 2-band model allows for 
analytic expressions for the frequencies $\omega_{\VEC{q}s}$ 
and polarization vectors (eigenmodes) $\epsilon_{\VEC{q}s}^j$, which will aid in the simplification and interpretation of the \Gru{} parameter and thermal conductivity computations. A 2-band dynamical matrix can be written as a $2\times 2$ matrix:
\begin{equation} K(\VEC{q})=\left(\begin{matrix}
 & d_0+d_3 & d_1-i d_2  \\
 & d_1+id_2 & d_0-d_3 \\
\end{matrix}\right)=d_0\mathit{I}+ \VEC{d} \cdot \VEC{\sigma}\end{equation}
where $d_0$ and $\VEC d=(d_1,d_2,d_3)$ are functions of $\VEC{q}$, and $\VEC \sigma$ is the vector of the Pauli matrices. In the presence of sublattice symmetry, as in the case of graphene, $d_3=0$. The eigenvalues $\omega_{\VEC{q}s}^2$ of $K(\VEC{q})$ satisfy (with $d=|\VEC d|=\sqrt{d_1^2+d_2^2+d_3^2}$)
\begin{equation}
\omega_{\VEC{q},\pm}=\sqrt{d_0\pm  d}
\end{equation}
with polarization vectors (eigenvectors) given by 
\begin{equation} 
\epsilon_\pm(\VEC{q})=\frac1{\sqrt{2d(d\pm d_3)}}\left(\begin{matrix}
 & d_3\pm d   \\
 & d_1+id_2\\
\end{matrix}\right) \xrightarrow[d_3=0]{}\frac1{\sqrt{2}}\left(\begin{matrix}
 & \pm 1   \\
 & e^{i\phi}\\
\end{matrix}\right)
\label{epsilon}
\end{equation}
where $\tan\phi=d_2/d_1$. For graphene, sublattice symmetry restricts $d_3$ to zero, and $\epsilon_\pm$ contains only the information on the winding of the vector $\VEC d$ in the $d_1$-$d_2$ plane. 

The explicit forms of $d_0,d_1,d_2$ and $d_3$ for the out-of-plane subspace of graphene are presented in Appendix \ref{sec:dvector_graphene}. The interested reader may also refer to Appendix \ref{sec:dvector} for an instructive derivation of the these quantities in a simple monoatomic unit cell lattice with non-rigid springs. 

\subsection{\Gru{} parameters for the flexural modes}


From Eq. \ref{grun0}, the \Gru{} parameter $\gamma_{\VEC{q} s}$ of a flexural mode $s=\pm$ takes the form (with a slight change of notation)
\begin{eqnarray}
\gamma_{\VEC{q} s}(\T)&=& \frac1{2M\omega^2_{\VEC{q} s}}\sum_{j,j'}
[\epsilon_{\VEC{q}s}^{j}]^*\notag\\&&
\left[\sum_{l'l'',j''} e^{i\VEC{q} \cdot  \VEC R_{l'}}\left(\VEC\Psi^{zz}  
\cdot (\T\VEC r)\right)_{0j,l'j',l''j''}\right]
\epsilon_{\VEC{q}s}^{j'}\notag\\
\label{gruneisenzz}
\end{eqnarray}
where\footnote{There is no $\Psi^{zzz}$ component due to antisymmetry.} $\VEC \Psi^{zz}=(\Psi^{zzx},\Psi^{zzy})$, and $\VEC r=(r^x,r^y)$ is the position of the $(j'',l'')$th atom. $M$ is the mass of the carbon atom. Notice that $\gamma_{\VEC{q} s}$ only depends on the CFCs of indices $\Psi^{zz\gamma}$, $\gamma=x,y$, since the flexural modes $\epsilon^{z,j}_{\VEC{q}s}=\epsilon^j_{\VEC{q}s}$ of our model (Eq. \ref{Force2}) lives in the $z$-polarization subspace. 

Under a crystal symmetry transformation ($C_3$ rotation, reflection) about $\VEC R_{l'}$, the (isotropic) \Gru{} parameter $\gamma_{\VEC{q}s}$ should remain invariant. Hence\footnote{Mathematically, a real 1-dimensional real representation of the dihedral group plus translation can only be the trivial representation}, all CFC terms related by crystal symmetry should have identical $\VEC\Psi^{zz} \cdot\VEC r=\Psi_0$. This implies that  $\VEC \Psi^{zz}=\Psi_0\VEC r/|\VEC r|^2$. 
With that, the anisotropic \Gru{} parameter with $\T = \VEC{\tf} \VEC{\tf}^T$ where $\VEC{\tf}  = (\cos \thetat, \sin \thetat, 0)$ simplifies to
\begin{eqnarray}
\gamma_{\VEC{q} s}(\T)&=&\frac1{2M\omega^2_{\VEC{q} s}}
\sum_{j,j'}[\epsilon_{\VEC{q}s}^{j}]^*\epsilon_{\VEC{q}s}^{j'}\notag\\&&
\times\left[\sum_{l'l'',j''} e^{i\VEC{q} \cdot  \VEC R_{l'}}\left[(\VEC\Psi^{zz}  \cdot\VEC{\tf} )(\VEC r\cdot \VEC{\tf})\right]_{0j, l'j',l''j''}\right]\notag\\
&=&\frac1{2M\omega^2_{\VEC{q} s}}
\sum_{j,j'}[\epsilon_{\VEC{q}s}^{j}]^*\epsilon_{\VEC{q}s}^{j'}\notag\\&&
\times\left[\sum_{l'l'',j''} e^{i\VEC{q} \cdot  
\VEC R_{l'}}[\Psi_0]_{0j,l'j',l''j''}(\hat{\VEC{r}}_{l''j''}\cdot \VEC{\tf})^2\right]
\label{gruneisenzza}
\end{eqnarray}
In other words, each type of CFC is characterized by a single constant $\Psi_0$, as well as 
a positive factor $(\hat{\VEC{r}}_{l''j''}\cdot \VEC{\tf})^2$ 
that depends on the directionality of the third (double-primed) atom from an arbitrary fixed origin. 

\subsection{CFCs of flexural graphene from the \Gru{} parameters}

We are now ready to demonstrate the central theme of this work, which is to reverse-engineer (see Sect. \ref{sec:reverse}) the coefficients of the CFCs from the \Gru{} parameters computed via DFT. Here the main focus is to provide a pedagogical demonstration of the physical insights gained from our approach, since the CFCs of graphene are not too numerically expensive to compute via other means. It must be stressed, however, that our reverse-engineering approach can be routinely generalized to materials with much more complicated symmetry based on the steps outlined at the end of Section \ref{sec:reverse}. 

\begin{figure}[htb]
\includegraphics[scale=.106]{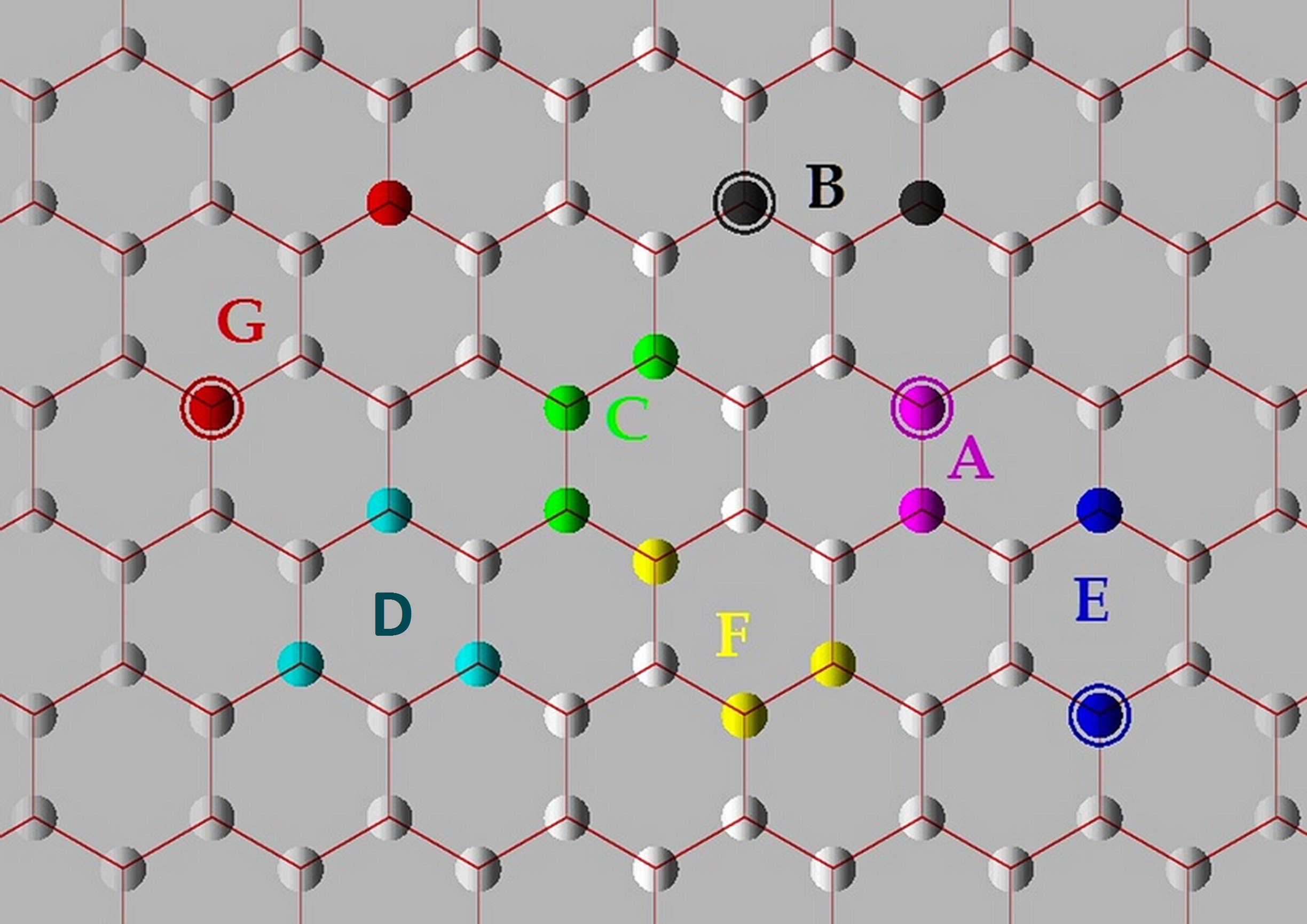}
\includegraphics[scale=.176]{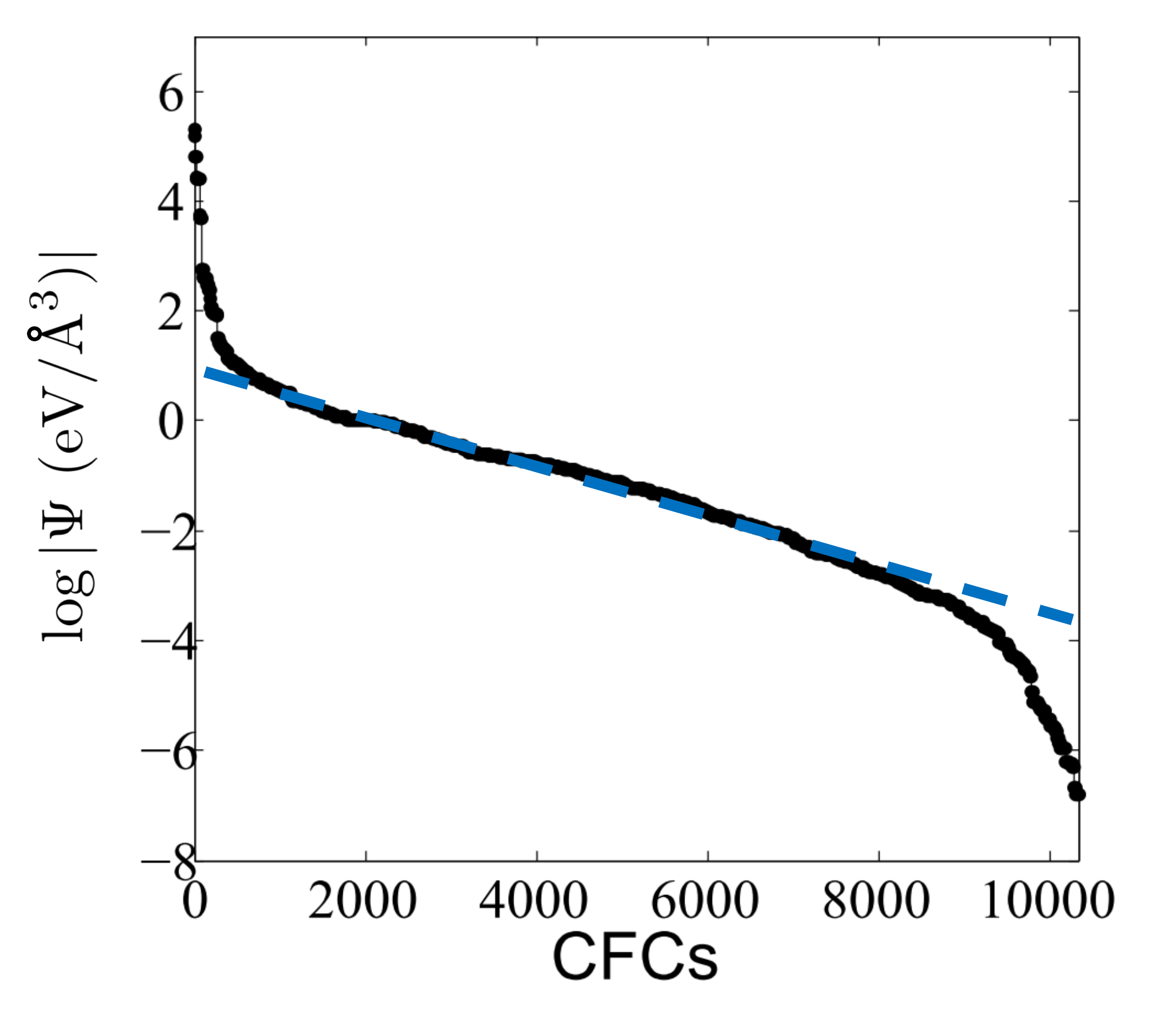}
\caption{(Color online) Left) CFC types $A$ to $G$ on a graphene sheet. $A,B$ and $G$ contains a singly occupied site and a doubly occupied site, while $C,D,E,F$ have all of their three sites singly occupied. Types $E$ and $F$ involve third nearest neighbors, while type $G$ involve a fourth nearest neighbor. Only one copy of each type is shown here; the rest are related by rotation, reflection or translation symmetry. Right) Logplot of magnitudes of all the CFC coefficients in graphene, according to data from Ref. \onlinecite{Lindsay14v89}. Beyond the several dominant types of terms, the remaining IFCs are much smaller and distributed exponentially (dashed blue guiding line). }
\label{grapheneIFCplot}
\end{figure}

We consider seven types of CFCs $A,B,...,G$
involving at most fourth nearest neighbors on the honeycomb lattice, as shown in Fig \ref{grapheneIFCplot}. 
By carefully keeping track of the all the terms related by symmetry, as well as their relative signs, the uniaxial (anisotropic) \Gru{} parameter can be expressed (via Eq. \ref{gruneisenzza}) in the form
\begin{equation}
\gamma_{\VEC{q}s}(\T)=\frac1{2M\omega_{\VEC{q} s}^2}\sum_{jj'}[\epsilon^j_{\VEC{q}s}]^*  \Gamma^{jj'}(\T)\epsilon_{\VEC{q}s}^{j'} 
\label{gamma}
\end{equation}
where $\Gamma(\T)=\Gamma_A+\Gamma_B...+\Gamma_G$ is a 
sum of $2\times 2$ matrices $\Gamma_A,...,\Gamma_G$ containing 
the contributions from the seven types of CFCs. 
The explicit forms of these matrices are presented in Appendix \ref{sec:graphene}. From 
Eq. \ref{gruneisenzza},  we obtain a few simplifying relations \footnote{This can be shown by exchanging the sublattice indices $j$ and $j'$, and simultaneously switching the origin of the vector $\hat{\VEC{r}}_{j''l''}$.} $\Gamma^{11}=(\Gamma^{22})^*$ and $\Gamma^{12}=(\Gamma^{21})^*$. Hence $\Gamma$ can also be decomposed in terms of the Pauli matrices 
via $\Gamma=\tau_0 \,\mathit{I}+\VEC \tau\cdot \VEC \sigma$ where $\VEC \tau = (\tau_1,\tau_2,i\tau_3)$\footnote{Note the factor of $i$ in the third component; in general, $\Gamma$ does not have to be Hermitian.}. The uniaxial \Gru{} parameter simplifies to
\begin{eqnarray}
\gamma_{\VEC{q}\pm }(\T)
&=&\frac1{2M\omega_{\VEC{q} \pm }^2}[\epsilon_{\VEC{q}\pm}]^* \cdot \Gamma \cdot \epsilon_{\VEC{q}\pm}\notag\\
&=&\frac1{2M\omega_{\VEC{q} \pm }^2}[\epsilon_{\VEC{q}\pm}]^* \cdot (\tau_0 \mathit{I}+\VEC \tau\cdot \VEC{\sigma}) \cdot \epsilon_{\VEC{q}\pm} \notag\\
&=&\frac1{2M\omega_{\VEC{q} \pm }^2}\left(\tau_0(\VEC{q}) \pm \VEC \tau(\VEC{q}) \cdot \hat{\VEC{d}}(\VEC{q}) \right)
\label{gamma2}
\end{eqnarray}
where $\hat{\VEC{d}} = \frac{1}{\sqrt{d_1^2+d_2^2}} (d_1,d_2)$, $d_1$ and $d_2$ being explicitly given in Eqs.~\ref{dvec1} and \ref{dvec2} for the tight-binding model given above. In our subsequent numerical computations, we shall however directly use polarization vector data from DFT. Note that $d_0$ cannot affect the polarization vectors, and do not enter Eq. \ref{gamma2} at all. 
Equation \ref{gamma2} elegantly expresses the \Gru{} parameter in terms of the 
``vectors" $\hat{\VEC{d}}$ and $\VEC{\tau}$ characterizing the 
quadratic and cubic IFCs, respectively, and is in fact applicable to a generic 2-band phonon model. Recalling that the physical interpretation of 
$\gamma_{\VEC{q}s}(\T)$ as the fractional change in $\omega_{\VEC{q}s} $ due to a fractional deformation in the direction of $\hat{\VEC{\tf}}$, we see that the change of $\omega^2_{\VEC{q}s}$ depends not just on the magnitude of the CFCs per se (as in a monoatomic lattice, see Appendix \ref{sec:graphene}), but also on their relative alignment with the quadratic IFCs in sublattice space. Specifically,
\begin{enumerate}
\item The contribution $\tau_0=\frac{\Gamma_{11}+\Gamma_{22}}{2}$ exists \emph{independently} from the phonon dispersion. It has equal weight in both sublattices, and couples to the (constant) normalization of $\epsilon_\pm$. 
\item The contribution $\VEC{\tau}\cdot \hat{\VEC{d}}$ depends on the phonon 
dispersion via $\hat{\VEC{d}}(\VEC{q})$. It arises from the couplings between the inter-sublattice 
components of $\Phi$ and $\Psi (\hat{\VEC{r}}\cdot \hat{\VEC{\tf}})^2$, which represent the quadratic and cubic IFCs, respectively. 
\end{enumerate}

\begin{figure}
\includegraphics[scale=.39]{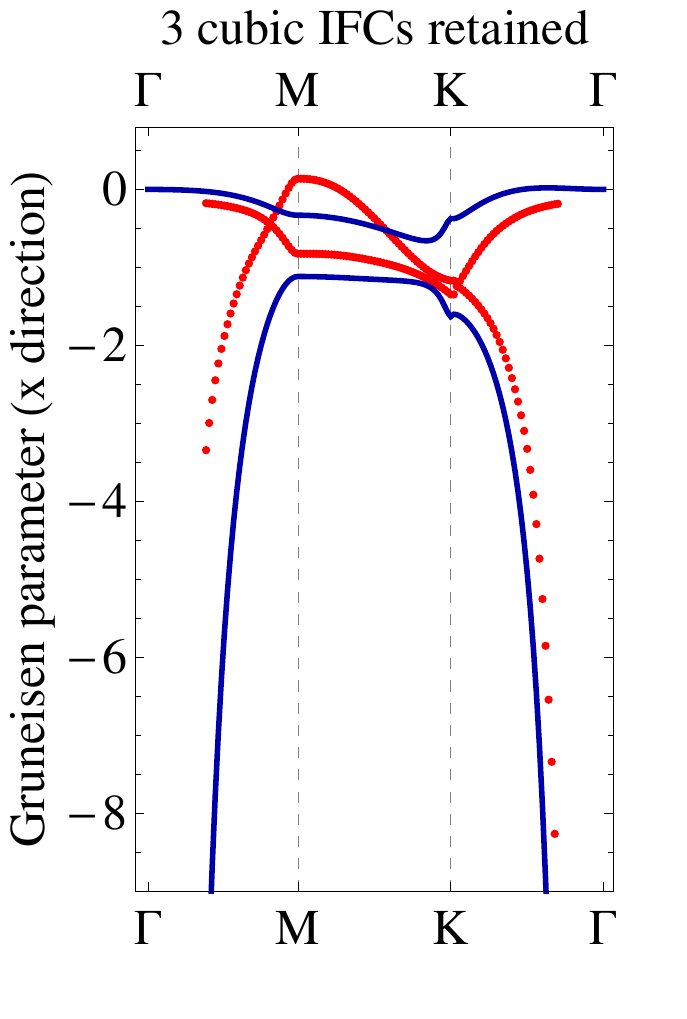}
\includegraphics[scale=.39]{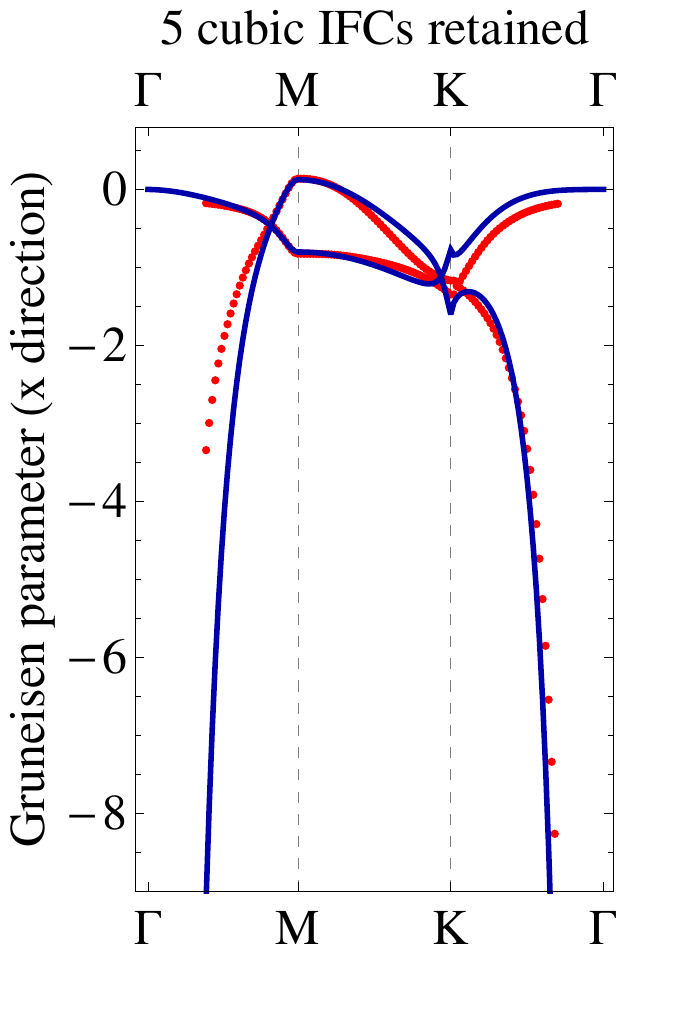}
\includegraphics[scale=.39]{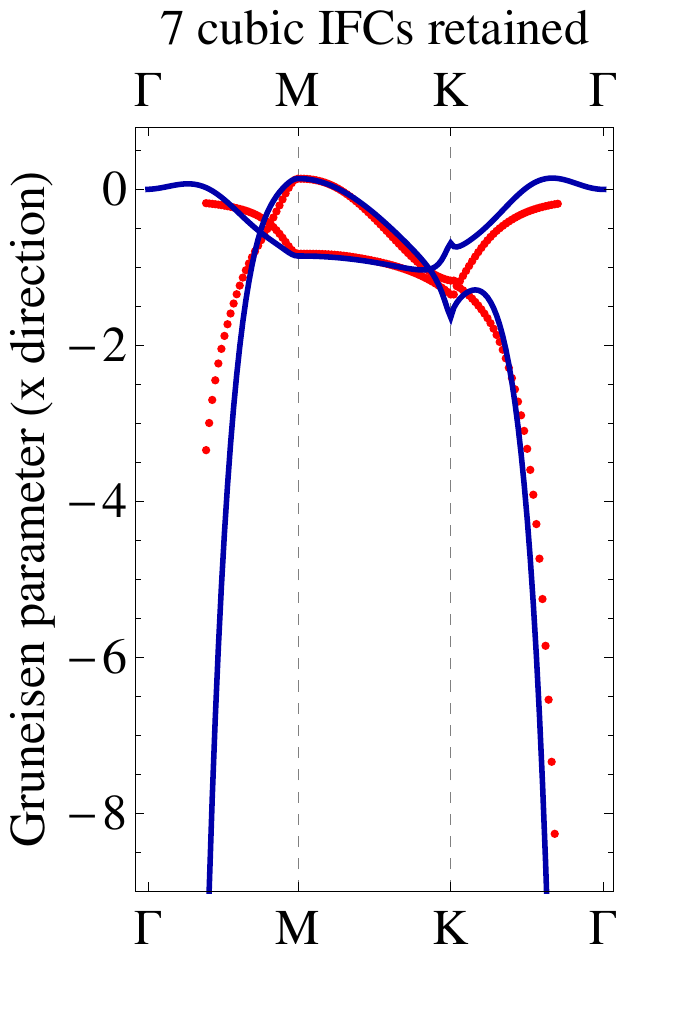}
\includegraphics[scale=.39]{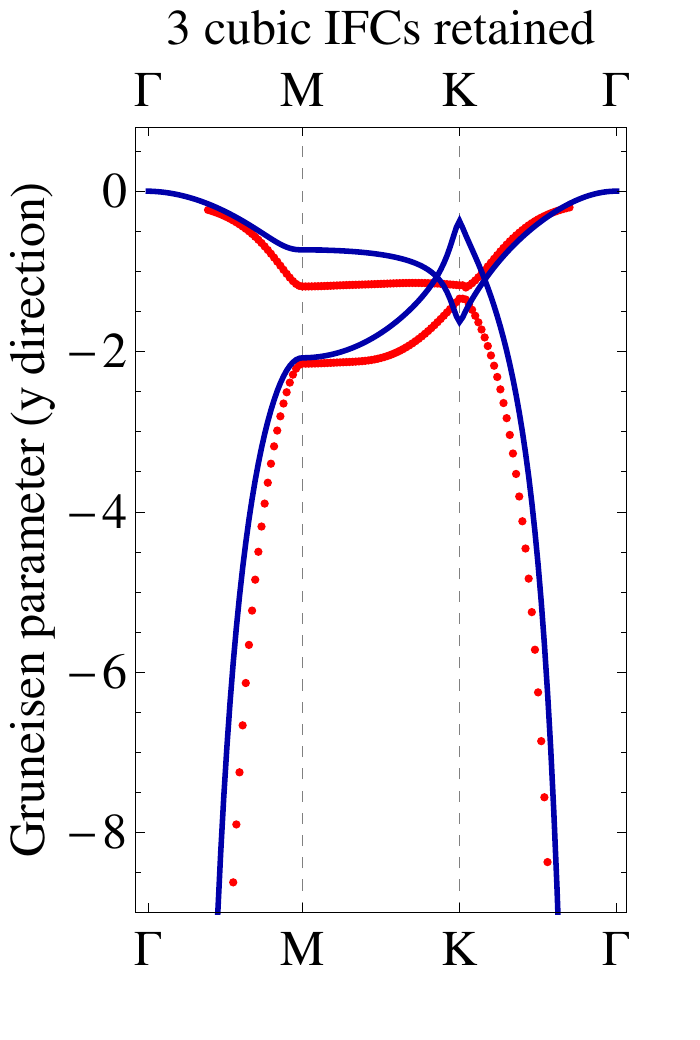}
\includegraphics[scale=.39]{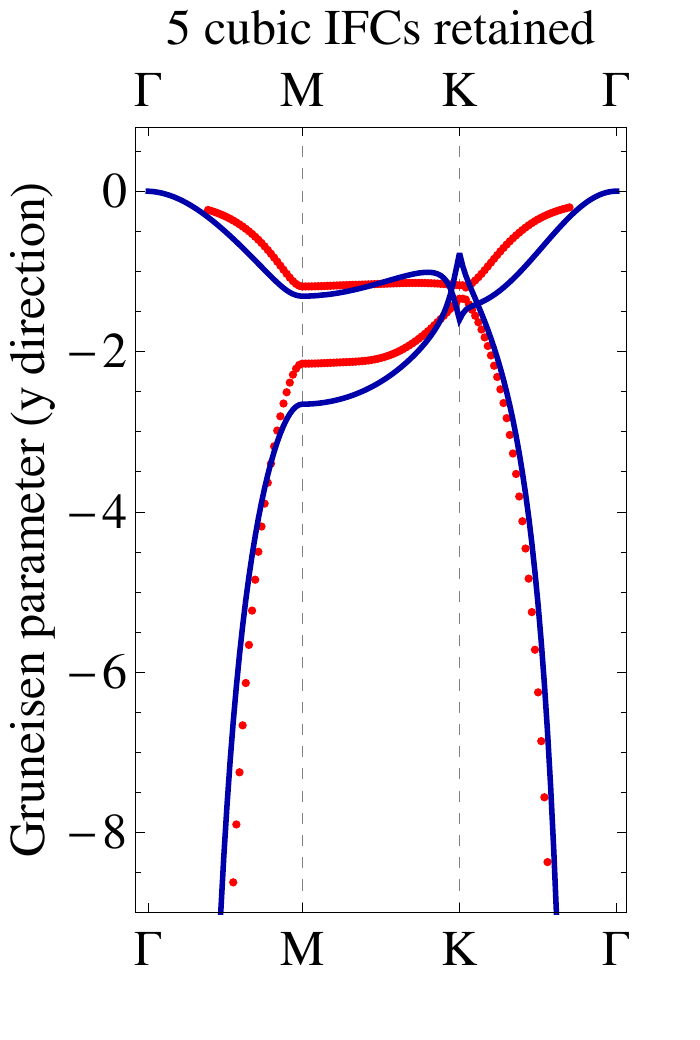}
\includegraphics[scale=.39]{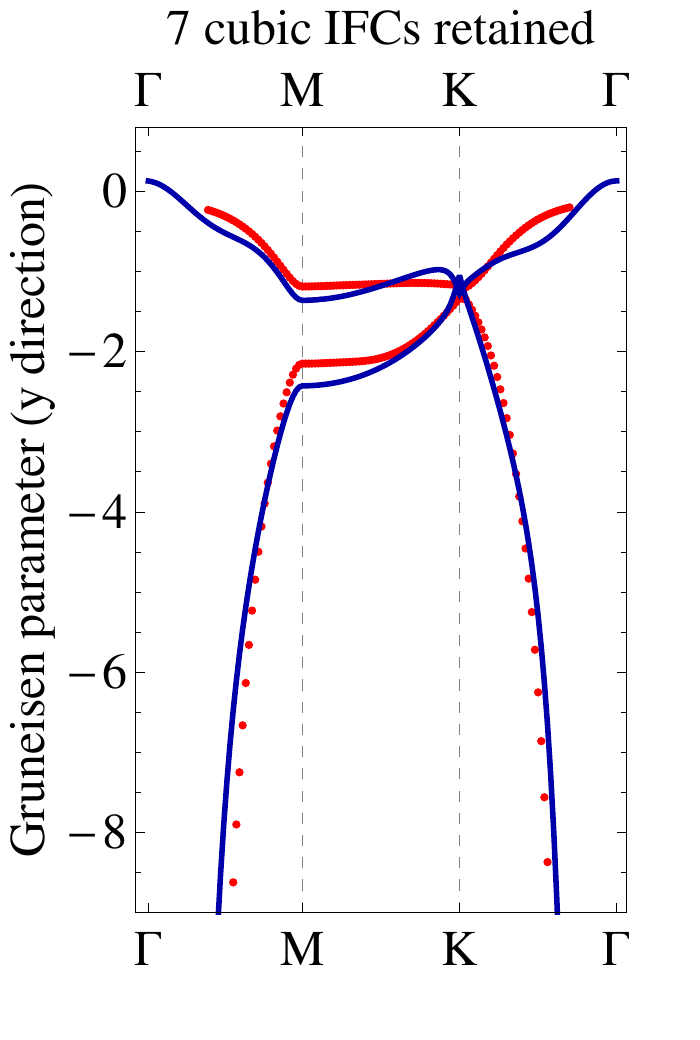}
\caption{(Color online) 
Comparison between the anisotropic \Gru{} parameters $\gamma_{\VEC{q}s}(\hat x),\gamma_{\VEC{q}s}(\hat y)$ obtained directly via DFT (red dotted curve), with those calculated from reverse-engineered CFCs via Eq. \ref{gamma2} (blue continuous curves), as described in the main text. 
Their agreement increases significantly as more CFCs are retained: The left,center and right columns depict cases with $\{ A,B,C\}$ types, $\{A,B,C,D,E\}$ types, and all seven CFC types $A$ to $G$, respectively.}
\label{gru}
\end{figure}

\begin{figure}
\includegraphics[scale=.2]{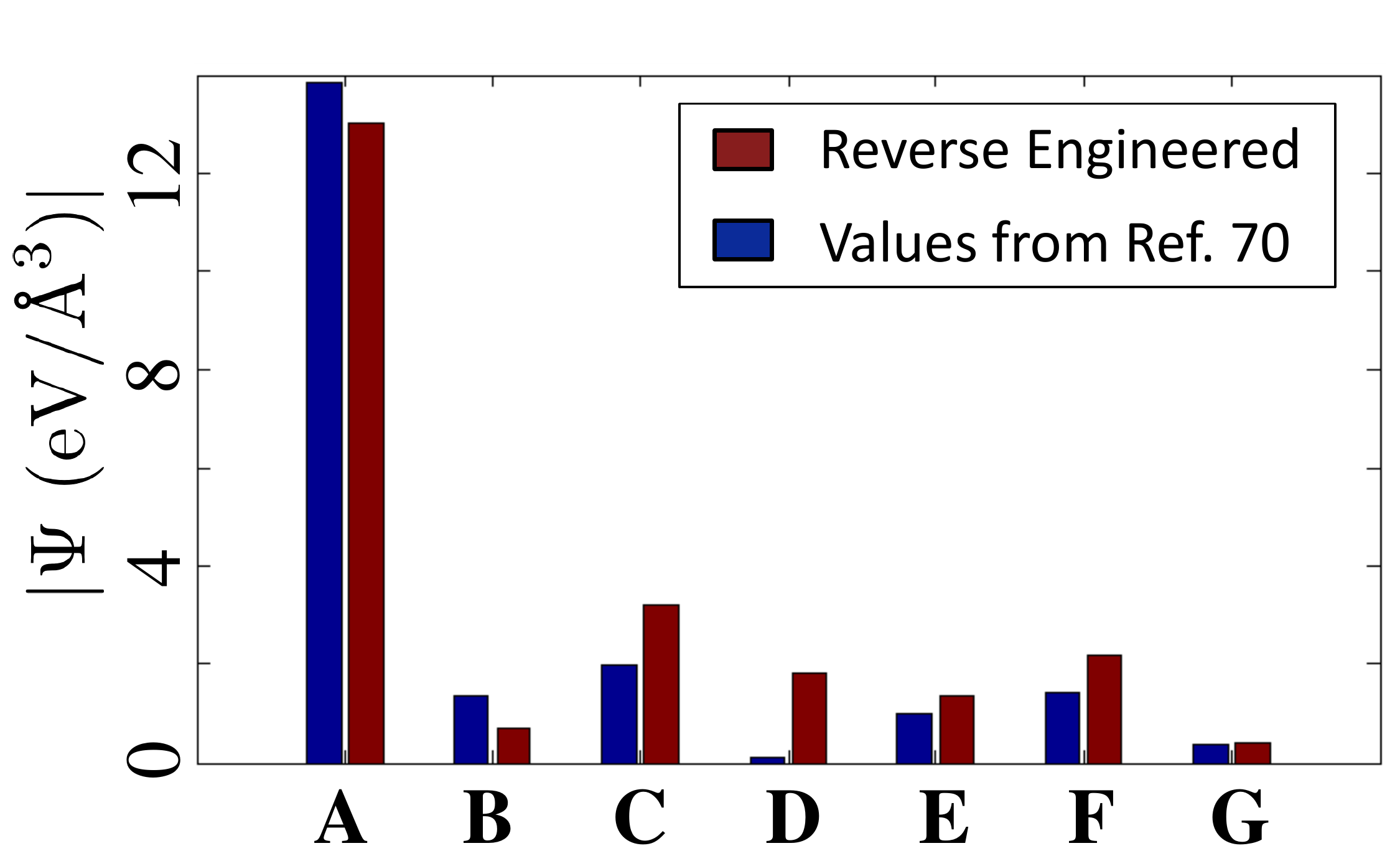}
\includegraphics[scale=.52]{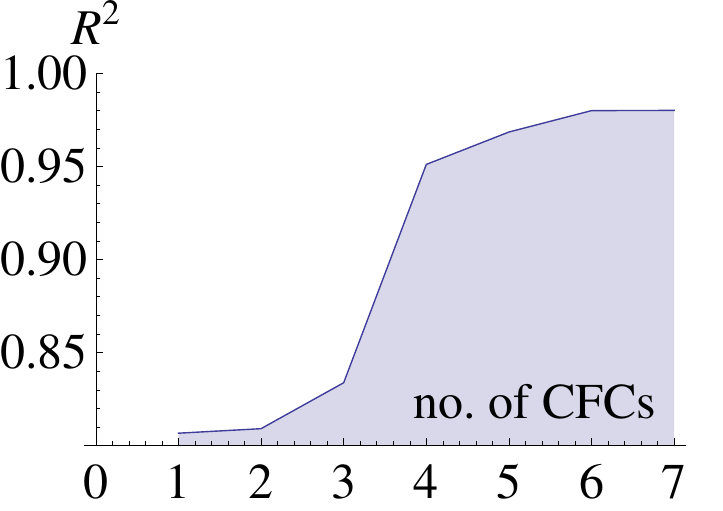}
\caption{(Color online) (Left) The relative magnitudes of the CFCs of types $A$ to $G$ we reversed engineered from our DFT data. The NN type $A$ dominates, but the NNN type $B$ has almost zero weight. Of also little importance is the NNNNN type $G$. Note the good agreement with established data from Ref. \onlinecite{Lindsay14v89}, at least for the more dominant CFCs. (Right) The agreement of fit of Fig 3, i.e. the comparison of the Gruneisen parameters directly obtained from DFT, versus those computed from the reverse engineered CFCs via Eq. \ref{Force2}. This agreement, which is quantified by the $R^2$ score, is plotted for different numbers of nonzero CFCs types $A$ through $G$. It improves significantly after adding the $D$-type CFC to types $A$ to $C$.}
\label{gru7}
\end{figure}

In Fig. \ref{gru}, we demonstrate the fidelity of the reverse-engineering
of the CFCs from DFT data, as described in Sect. \ref{sec:reverse}. The
uniaxial \Gru{} parameters $\gamma_{\VEC{q}\pm}(\hat x)$ and
$\gamma_{\VEC{q}\pm}(\hat y)$ are first computed via DFT (red dotted curves).
We note that under an uniaxial deformation, the crystal type changes from hexagonal to
orthorhombic base-centered and the total CPU time for one set of phonon calculations (for a complete
phonon dispersion with 73 irreducible $k$ points) is typically less than
50 cpu hours on a Intel(R) Xeon(R) $2.30$~GHz multi-CPU workstation.
By substituting the polarization vectors in Eq. \ref{rev2} with the polarization eigenvectors obtained from DFT ,
the \Gru{} parameters was used to obtain the coefficients of seven shortest range nonzero CFCs of the forms $A,B,...,G$, as described in Fig. \ref{grapheneIFCplot} and numerically presented in fig. \ref{gru7}. In this simple example of flexural graphene, Eq. \ref{rev2} reduces to Eq. \ref{gamma2} had tight-binding parameters been used for the harmonic part of the phonon potential. 


The truncation of the CFCs leads to an incomplete basis, and we cannot expect to obtain perfectly accurate reverse-engineered CFC coefficients. To investigate how closely they resemble the true CFC coefficients, we compare their corresponding uniaxial \Gru{} parameter curves with those obtained from DFT in Figs. \ref{gru} and \ref{gru7}. We observe rather good agreement (with goodness-of-fit parameter $R^2>0.95$) upon the inclusion of $4$ or more CFCs, and even better agreement ($R^2>0.98$) with all seven types of CFCs $A$ to $G$. Explicitly, the individual CFCs reversed engineered from our DFT data also exhibit rather good agreement with established results\cite{Lindsay14v89}, especially for types A,B,C,E and G. 

This overall reasonable agreement suggests the usefulness of our reverse-engineering approach: with the \Gru{} parameters obtained from just a few DFT computations (one for each direction), we can already generate the largest CFCs rather accurately. This is contrasted with conventional, more computationally expensive DFPT methods that obtain the full set of CFCs through individual lattice perturbations. Note the crucial role of the directionality of the anisotropic \Gru{} parameters in isolating the CFCs: had the reverse engineering been performed with the (usual) \emph{isotropic} \Gru{} parameter instead, only a certain linear combination of the CFCs could have been extracted, as elaborated at the end of Appendix \ref{sec:graphene}. 

\subsection{Thermal conductivity of flexural modes}
\label{flexuralcond}
Finally, we shall illustrate the physical correctness of our CFC results by using them to calculate the thermal conductivity contribution from the acoustic flexural mode of graphene. This mode provides the largest contribution to the conductivity of graphene at room temperature or below\cite{Paulatto13v87}, and can indeed be estimated fairly accurately by our approach.

The peculiar quadratic dispersion of the acoustic flexural mode $Z_A$ places selection rules on possible three phonon scattering processes from the $Z_A$ channel\cite{bonini2007phonon}. Specifically, only processes involving two $Z_A$ phonons and one transverse or longitudinal in-plane ($P_A$) acoustic phonon are permitted\cite{lindsay2010flexural}. As such, the phonon broadening width (Eq. \ref{tau2}), which is integral to computing the conductivity, can be simplified to  
\begin{eqnarray}
\tau_{\VEC{q}, Z_A}^{-1}&=& \frac{\pi\hbar}{4N}\sum_{\VEC{q}'}|M_{\VEC{q},Z_A,\VEC{q}',Z_A,-\VEC{q}-\VEC{q'},P_A}|^2\notag\\
&&\times (n_{\VEC{q}',Z_A}-n_{-\VEC{q}-\VEC{q}',P_A})\delta_{\omega_{\VEC{q}',Z_A}+\omega_{\VEC{q},Z_A}-\omega_{-\VEC{q}-\VEC{q}',P_A}}.\notag\\
\label{tau4}
\end{eqnarray} 

To proceed, the matrix elements $M_{\VEC{q},Z_A,\VEC{q}',Z_A,-\VEC{q}-\VEC{q'},P_A}$ are evaluated in terms of our reverse-engineered CFCs via Eq. \ref{tau3}.  The polarization vectors $\epsilon_{\VEC{q},P_A}$ and $\epsilon_{\VEC{q},Z_A}$ in Eq. \ref{tau3} can be obtained by diagonalizing the stiffness matrix in Eq. \ref{Force2}. 
\begin{figure}
\includegraphics[scale=.7]{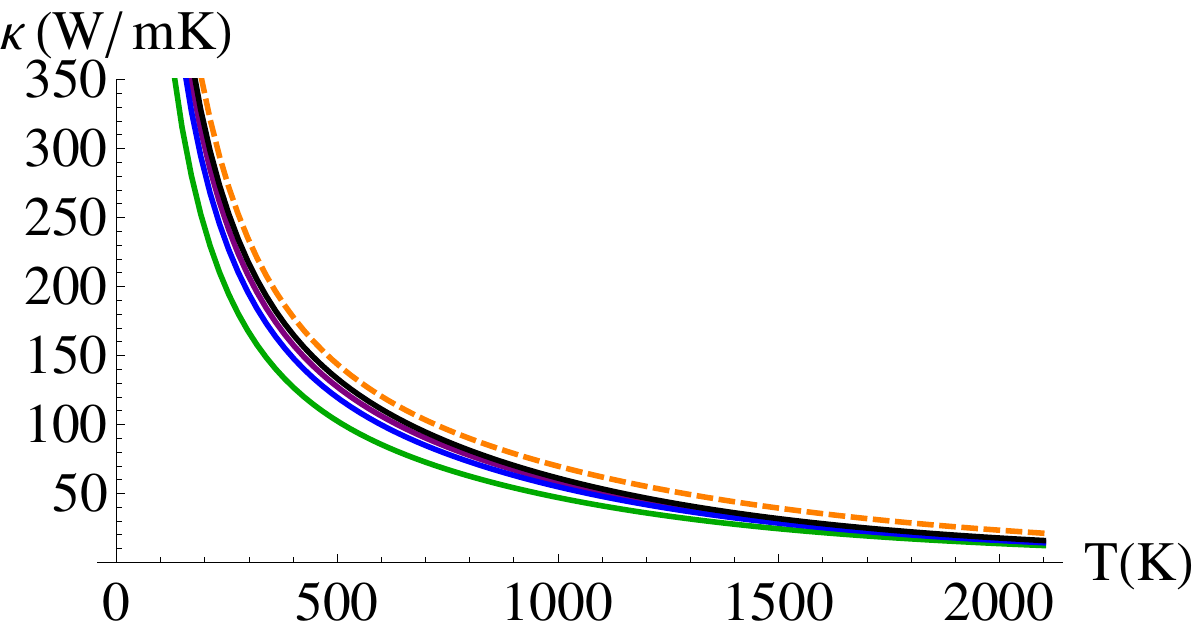}
\caption{(Color online) The thermal conductivity due to the $Z_A$ flexural mode of graphene, plotted as a function of temperature. Plotted are the contributions from only the A-type CFC (Green), as well as contributions from the closest 5 (Blue), 6 (Purple) and 7 (Black) CFCs. 
We observe a convergence towards established results (orange dashed curve) computed with an extremely fine mesh\cite{Paulatto13v87}, with the slight underestimate of less than $10\%$ attributable to the small number of CFCs we included. All CFCs were reverse-engineered from \Gru{} parameter data via Eq. \ref{rev2}, with polarization vectors obtained from DFT.}
\label{fig:cond}
\end{figure}
Notably, the thermal conductivity $\kappa$ diverges as $\sim T^{-1}$ at low temperatures. This can be understood from the small $\VEC{q}$ behavior of $\tau_{\VEC{q}, Z_A}^{-1}$. Since the $P_A$ modes are linearly dispersive with much steeper slopes than that of $Z_A$ (see Fig. \ref{bandfitting}), we necessarily require $\VEC{q}\approx -\VEC{q}'$ for simultaneous satisfaction of momentum and energy conservation. This leads to further simplification of Eq. \ref{tau4} via  $n_{\VEC{q}',Z_A}-n_{-\VEC{q}-\VEC{q}',P_A}\approx (2\sinh x_{\VEC{q},Z_A})^{-1}$, $x_{\VEC{q},Z_A}=\frac{\hbar \omega_{\VEC{q},Z_A}}{2k_BT}$. With that, one deduces that $\tau_{\VEC{q}, Z_A}^{-1}$ vanishes quadratically around the $\Gamma$ point, in agreement with results from Ref. \onlinecite{Paulatto13v87}. Together with the approximate $(2\sinh x_{\VEC{q},Z_A})^{-1}$ factor from the occupation functions, the $\kappa\sim T^{-1}$ behavior follows from simple Taylor expansion.

As demonstrated in Fig. \ref{fig:cond} our results deviate from that of a much more exhaustive computation (Ref. \onlinecite{Paulatto13v87}) by not more than $10\%$, a reasonable margin given that we have included contributions from only the seven types of CFC mentioned. Due to the smallness of the CFCs beyond the fifth, we observe a nice convergence pattern when more than 5 CFCs are included. Larger values of $\kappa$ from both ours and Ref. \onlinecite{Paulatto13v87}'s results were reported by other works i.e. Ref.~\onlinecite{Lindsay14v89}. This discrepancy is most probably due to the relaxation-time approximation, which may not be very accurate for graphene. Secondarily, we have also excluded  contributions from phonon-isotope and boundary scattering, both of which are beyond the scope of this study on anharmonicity.

\section{Discussion and Conclusion}
 
We have proposed and demonstrated a general approach for determining
the most important third-order interatomic force constants (also called the cubic
force constants (CFCs)) from \Gru{}
parameter data corresponding to deformations in appropriately chosen
directions. This approach can be viewed as a hybrid between real and reciprocal space approaches, where a derivative corresponding to a real-space deformation is applied to the dynamical matrix defined in reciprocal space. The high efficiency of our approach critically hinges upon the optimal use of density-functional perturbation
theory (DFPT) in phonon calculations, where a primitive cell is all that is required for deriving the phonon dispersions. 
To facilitate the retrieval of the CFCs, we have derived a most general expression for the \Gru{} parameters corresponding to very general deformation on all (seven) crystal types. 

That our approach even works appears counterintuitive at first sight, since \Gru{} parameters are defined with respect to uniform crystal distortions while the CFCs obviously have to be probed with independent individual atomic displacements. Yet, by virtue of the locality of typical atomic orbitals, we showed that \Gru{} data can indeed be used to ``reverse engineer" the values of a small but very important fraction of the CFCs. Testimony to the reliability of our approach is the excellent
fit of our ``reverse engineered" CFCs of graphene to well-established
results on the \Gru{} parameter, which also results in reasonably accurate
predictions of the conductivity due to the flexural channel.
It is expected that with systematic computer implementation of the general approach outlined in this paper, anharmonic effects such as the thermal conductivity of technologically important materials can be efficiently and routinely predicted with a first-principles treatment.

\section{Acknowledgments}

We gratefully acknowledge the harmonic and anharmonic force constants provided by Lucas~Lindsay of Oak Ridge National Laboratory.

\bibliographystyle{apsrev4-1}
\bibliography{references,ckgref} 

\appendix

\section{Derivation of thermal expansion in terms of \Gru{} parameters}
\label{sec:TECgru}
Here we derive from first principles Eq. \ref{TEC2}, which relates the thermal expansion coefficient (TEC) $\alpha$ in terms of the phonon mode heat capacities and \Gru{} parameters. $\alpha$ is defined as the fractional change in volume $V$ with temperature $T$  measured at pressure $P=0$:
\begin{equation}
\alpha = \left.\frac1{V}\frac{\partial V}{\partial T}\right|_{P=0}
\label{TEC}
\end{equation}
To derive its explicit dependence on the \Gru{} parameters and hence the CFCs, we first note that the condition $P=0$, where $\alpha$ is defined, 
constraints the lattice free energy $F$ via
\begin{equation}
P=-  \left.\frac{\partial F}{\partial V}\right|_T=-\left.\frac{\partial F_{\rm PE}}{\partial V}\right|_T-\left.\frac{\partial F_{\rm KE}}{\partial V}\right|_T=0
\label{P}
\end{equation}
where $F_{\rm PE}$ and $F_{\rm KE}$ are the potential and kinetic portions of the free energy. The first term is related to $\Delta V$, the volume change due to increasing temperature 
via $\left.\frac{\partial F_{\rm PE}}{\partial V}\right|_T=C\Delta V$, which is proportional to the elastic constant matrix $C$. The second term $F_{\rm KE}=\hbar\sum_{\VEC{q}s}\omega_{\VEC{q}s}$ is a sum over the mode energies, so $\left.\frac{\partial F_{\rm KE}}{\partial V}\right|_T=\hbar\sum_{\VEC{q} s}\frac{\partial F_{\rm KE}}{\omega_{\VEC{q} s}}\frac{\partial \omega_{\VEC{q} s}}{\partial V}=-\sum_{\VEC{q} s}\frac{V\bar \epsilon_{\VEC{q} s}}{\omega_{\VEC{q} s}}\frac{\partial \omega_{\VEC{q} s}}{\partial V}=\sum_{\VEC{q} s}\gamma_{\VEC{q} s}\bar \epsilon_{\VEC{q} s}$ where $\bar\epsilon_{\VEC{q} s}$ is the equilibrium energy of the phonon mode. Combining the above in Eqs. \ref{TEC} and \ref{P} and employing the tensorial properties of the derivatives, we obtain Eq. \ref{TEC2}:
\begin{equation}
\alpha_i = \frac1{\Omega}C^{-1}_{ij}I_j,
\end{equation}
\begin{equation}
I_j=\frac{\Omega}{V}\sum_{\VEC{q} s} \gamma_{\VEC{q} s} c_{\VEC{q} s},
\end{equation}
where $ c_{\VEC{q} s}=\left.\frac{\partial\bar \epsilon_{\VEC{q}s}}{\partial T} \right|_V$ is the phonon mode heat capacity, and $\Omega$ 
the equilibrium unit cell volume.

\section{Spring toy model}

To provide a pedagogical and explicit illustration of how the CFCs contribute to the \Gru{} parameter, we consider a toy model of a rectangular lattice of anharmonic springs. Denote the length, stiffness and direction vector of each spring as $a$, $k$ and $\hat{\VEC{n}}$. A spring between lattice sites $l$ and $l+1$ contributes a potential energy of 
\begin{eqnarray}
 U_l &= &\frac{k}{2}(\hat{\VEC{n}}\cdot \Delta \VEC u)^2 + \frac{\lambda}{6}(\hat{\VEC{n}}\cdot \Delta \VEC u)^3\notag\\
&=& \frac{k}{2}\sum_{\alpha\beta}\Delta u^\alpha[\hat{\VEC{n}} \hat{\VEC{n}}^T]^{\alpha\beta}\Delta u^\beta +\frac{\lambda}{6}\sum_{\alpha\beta\gamma}\hat{\VEC{n}}^\alpha\hat{\VEC{n}}^\beta \hat{\VEC{n}}^\gamma \Delta u^\alpha \Delta u^\beta \Delta u^\gamma\notag\\
\label{Ui}
\end{eqnarray}
where $\Delta \VEC u=\VEC u_{l+1}-\VEC u_l$ is the difference between the displacements about the equilibrium positions at sites $l+1$ and $l$. These sites themselves are physically separated by a displacement $a\hat{\VEC{n}}$. By expanding $(\Delta \VEC u)^3=(\VEC u_{l+1}-\VEC u_l)^3$ and summing over the lattice, the cubic term in Eq. \ref{Ui} takes the form\footnote{Note that the CFCs corresponding to $u_l^3$ and $u_{l-1}^3$ annihilate each other.}
\begin{widetext}
\begin{eqnarray}
&&\frac1{6}\sum_{\alpha\beta\gamma}\sum_{l,l',l''}\Psi^{\alpha\beta\gamma}_{l,l',l''}u_l^\alpha u_{l'}^\beta u_{l''}^\gamma\notag\\
&=&\frac{\lambda}{6}\sum_{\alpha\beta\gamma}\left[\hat{\VEC{n}}^\alpha\hat{\VEC{n}}^\beta \hat{\VEC{n}}^\gamma \left(\sum_{l,l',l''} \delta_{l'l}(\delta_{l'',l+1}-\delta_{l'',l-1})+\delta_{l',l+1}(\delta_{l'',l}-\delta_{l'',l+1})+\delta_{l',l-1}(\delta_{l'',l-1}-\delta_{l'',l})\right)\right]u_l^\alpha u_{l'}^\beta u_{l''}^\gamma\notag\\
\end{eqnarray}
\end{widetext}
The quantity in the parentheses are just the six surviving cross  
 terms, given for instance by $-u_l^\alpha u_{l}^\beta u_{l-1}^\gamma \hat{\VEC{n}}^\alpha\hat{\VEC{n}}^\beta\hat{\VEC{n}}^\gamma$ or $u_l^\alpha u_{l-1}^\beta u_{l-1}^\gamma \hat{\VEC{n}}^\alpha\hat{\VEC{n}}^\beta\hat{\VEC{n}}^\gamma$. 

To compute the \Gru{} parameter, we substitute $a\hat{\VEC{n}} =a(\cos\theta,\sin\theta)^T$ for $r^\gamma_{j''l''}$ in Eq. \ref{grun0}. For the case of a trivial unit cell, the $j$ sublattice index is irrelevant and the quantity in the square parentheses of Eq. \ref{grun0} simplifies to
\begin{equation}
\sum_{l'} e^{i\VEC{q} \cdot  \VEC R_{l'}}\left(\sum_{l'',\gamma}\Psi^{\alpha\beta\gamma}_{0l'l''}r_{l''}^\gamma\right)= 2a\lambda (1-\cos[a\VEC{q} \cdot \hat{\VEC{n}} ] )\left [\hat{\VEC{n}} \hat{\VEC{n}}^T\right]^{\alpha\beta}
\end{equation}

We next make use of $[\epsilon_s^\alpha]^* \left[\hat{\VEC{n}} 
\hat{\VEC{n}}^T\right]^{\alpha\beta}
\epsilon_s^\beta=|\VEC{\epsilon}_s^*\cdot \hat{\VEC{n}}|^2$ to obtain, for a lattice with various spring types $i=1,2,...$, 
\begin{eqnarray}
\gamma_{\VEC{q}s}&=&\frac1{\Lambda M\omega^2_{\VEC{q},s}}\sum_{i}\lambda_ia_i(1-\cos[a_i\VEC{q} \cdot \hat{\VEC{n}}_i ] )|\VEC\epsilon_{\VEC{q}s}^*\cdot \hat{\VEC{n}}_i|^2\notag\\
\label{gammae}
\end{eqnarray}
where $M$ is the mass of each atom.
In the simplest case of a 1-dimensional lattice with a trivial unit cell, $\VEC\epsilon_s$ is trivial and $|\VEC\epsilon_{\VEC{q}s}^*\cdot \hat{\VEC{n}}_i|^2=1$. 

If the lattice is 2-dimensional, but still with a trivial unit cell, $\VEC\epsilon_s$ has two polarization components and we can simplify Eq. \ref{gammae} via Eq. \ref{gamma2} to obtain
\begin{eqnarray}
&=&\frac1{2\Lambda M\omega^2_{\VEC{q},\pm}}\sum_{i}\lambda_ia_i(1-\cos[a_i\VEC{q} \cdot \hat{\VEC{n}}_i ] )(1\pm \cos(2\theta_i-\varphi(\VEC{q})))
\notag\\
\end{eqnarray}
where $d_1=d\sin\varphi$, $d_3=d\cos \varphi$. The last line can also be obtained via the explicit form of $\epsilon_\pm$ from Eq. \ref{epsilon}. 

For a simplest example, consider the case with only one type of spring (i.e., a linear mono-atomic chain), $d_0(q)=d(q)$, $d_1(q)=d(q)\sin 2\theta$, $d_3(q)=d(q)\cos 2\theta$, so that we simply have $\varphi =2\theta$. 
This yields $\gamma_{q,+}=\frac{\lambda a}{2kD}=\frac{\lambda a}{2k}$ for the optical mode, in agreement with textbook calculations.\cite{Mihaly96-book}

\section{Determination of the dynamical matrices for a monoatomic and a diatomic system}
\subsection{Monoatomic lattice with non-rigid springs}
\label{sec:dvector}
To find the dynamical matrix (Eq. \ref{dynamical}), one takes the Fourier transform of the real space quadratic IFCs. In a monoatomic lattice, the sublattice index $j$ is irrelevant and we write $D(\VEC{q})=d_0\mathit{I}+\VEC d\cdot\VEC \sigma$, where $\VEC \sigma$ is the vector of Pauli matrices. Each non-rigid spring in the direction $\hat{\VEC{n}}$ can only exert a force of $-(\Delta \VEC u\cdot \hat{\VEC{n}})\hat{\VEC{n}}\propto \hat{\VEC{n}}$, where $\Delta\VEC u$ is the difference of the displacements $\VEC u$ of its two ends. 

In a monoatomic lattice, each atom is connected by equivalent springs in opposite directions, with extensions given by $\Delta \VEC u=\VEC u_{l+1}-\VEC u_l$ and $\Delta \VEC u=\VEC u_{l}-\VEC u_{l-1}$. Hence a pair of springs along the direction $\hat{\VEC{n}}=(\cos \theta, \sin \theta)^T$ contributes a factor of 
\begin{eqnarray}
&&\frac{k}{M}(2-e^{ia\VEC{q} \cdot \hat{\VEC{n}} }-e^{-ia\VEC{q} \cdot \hat{\VEC{n}} })(\hat{\VEC{n}}\hat{\VEC{n}}^T)\notag\\
&=&\frac{k}{M}(1-\cos[a\VEC{q} \cdot \hat{\VEC{n}} ])(\mathit{I}+\sigma_1 \sin 2\theta + \sigma_3 \cos 2\theta )\notag\\
\end{eqnarray}
to $D(\VEC{q})$, where $M$ is the mass of each atom, $a$ the spring length and $k$ the spring stiffness\footnote{This expression is manifestly symmetric under $\hat{\VEC{n}}\rightarrow -\hat{\VEC{n}}$, as it should be.}. If each atom is attached to springs of stiffness $k_1,k_2,...$, lengths $a_1,a_2,...$ oriented in directions $\hat{\VEC{n}}_i=(\cos \theta_i, \sin \theta_i)^T$, $i=1,2,...$, we have
\begin{subequations}
\begin{equation} d_0=\frac1{M}\sum_i k_i(1-\cos[a_i\VEC{q} \cdot \hat{\VEC{n}}_i ]) \end{equation}
\begin{equation} d_1=\frac1{M}\sum_i k_i(1-\cos[a_i\VEC{q} \cdot \hat{\VEC{n}}_i ])\sin 2\theta_i \end{equation}
\begin{equation} d_3=\frac1{M}\sum_i k_i(1-\cos[a_i\VEC{q} \cdot \hat{\VEC{n}}_i ])\cos 2\theta_i \end{equation}
\begin{equation}d_2=0\end{equation}
\end{subequations}
Note that with $\VEC d$ living in the space of phonon polarizations, as above (but not in the following subsection), we always have $d_2=0$ in the presence of time-reversal symmetry.

\subsection{Flexural modes on diatomic graphene lattice}
\label{sec:dvector_graphene}

In the case of flexural (out-of-plane) phonon modes in 
the force constant model of graphene (Eq. \ref{Force2}), we also have a decoupled two-band sector due to the decoupling of in-plane and out-of-plane modes. However, these two bands now represent sublattice DOFs associated with out-of-plane movements of atoms, rather than the $x$ and $y$ polarizations studied in Section|\ref{sec:dvector}.
By taking lattice Fourier transforms (see Fig. \ref{grapheneIFCplot}), we find 
$D(\VEC{q})= \frac1{M} \VEC{\sigma}\cdot \VEC d(\VEC{q})$ with $\VEC{q}=\frac{1}{a}(q_1,q_2)$, $a$ being the lattice constant, and
\begin{eqnarray}
d_0 & = & 3t_{1}+2t_{2}(3-\cos q_1-\cos q_2 - \notag
\\ & &  \cos(q_1+q_2))+3t_3+6t_4 
\\ d_1 &=& t_{1}(1+\cos q_1 + \cos(q_1+q_2)) + \notag
\\ & & t_3(2\cos q_2 +\cos(2q_1+q_2))+ \notag
\\ & & t_4(\cos q_1+\cos 2q_1 +2\cos q_1\cos q_2+ \notag
\\ & & \cos(q_1+2q_2)+\cos(2q_1+2q_2)) 
\label{dvec1}
\\d_2 &=& t_{1}(\sin q_1+\sin (q_1+q_2)) +t_3(\sin(2q_1+q_2))+ \notag
\\ &  & t_4(-\sin q_1+\sin 2q_1 +\sin(q_1-q_2)+\sin(q_1+q_2)+ \notag
\\& & \sin(q_1+2q_2)+\sin(2q_1+2q_2))  
\label{dvec2}
\\d_3 &=& 0
\end{eqnarray}
where $t_1,t_2,t_3,t_4$ are the first, second, third and fourth nearest-neighbor 
quadratic IFC coefficients given by Ref. \onlinecite{Zimmermann08v78}. 
Note that we now have $d_3=0$ due to sublattice symmetry, in contrast
to the spring systems treated in section~\ref{sec:dvector}, where $d_2=0$.

\section{\Gru{} parameters from graphene CFCs}
\label{sec:graphene}
Here, we present the detailed expressions for $\Gamma$ in the expression for $\gamma_{\VEC{q}s}(\T)$ (Eqs. \ref{gamma} and \ref{gamma2}). These results are based on the CFCs $A$ to $G$ in Fig. \ref{grapheneIFCplot}, whose exact configurations affect $\Gamma$ via Eq. \ref{Force2}.

As explained in the main text, we always have $\Gamma^{11}=(\Gamma^{22})^*$ and $\Gamma^{12}=(\Gamma^{21})^*$. Denoting $\VEC{\tf}=(\cos\thetat,\sin\thetat,0)$, we have
\begin{widetext}
\begin{align}
 \Gamma_A^{12}&= -\Psi_A \left(\cos^2\left(\frac{\pi}{2}-\thetat\right)e^{-i(q_1+q_2)}+ \cos^2\left(\frac{\pi}{6}-\thetat\right)+\cos^2\left(\frac{5\pi}{6}-\thetat\right)e^{-iq_1}\right)\\
 \Gamma_A^{11}&=\frac{3}{2}\Psi_A 
\\ \notag \\
\Gamma_B^{12}&=0\\
\Gamma_B^{11}&=\Psi_B \left(\frac{3}{2}- \cos^2\left(\thetat\right)e^{i q_1}-\cos^2\left(\frac{\pi}{3}-\thetat\right)e^{i(q_1+q_2)}-\cos^2\left(\frac{2\pi}{3}-\thetat\right)e^{i q_2}\right)
\\ \notag \\
\Gamma_C^{12}&= \Psi_C \left( \cos^2\left(\frac{\pi}{2}-\thetat\right)(e^{-iq_1}-1)+\cos^2\left(\frac{\pi}{6}-\thetat\right)(e^{-i(q_1+q_2)}-e^{-iq_1})+\cos^2\left(\frac{5\pi}{6}-\thetat\right)(1-e^{-i(q_1+q_2)})\right)\\
\Gamma_C^{11}&= \Psi_C \left( \cos^2\left(\frac{\pi}{2}-\thetat\right)(e^{-iq_2}-e^{-i(q_1+q_2)})+\cos^2\left(\frac{\pi}{6}-\thetat\right)(e^{i(q_1+q_2)}-e^{iq_1})+\cos^2\left(\frac{5\pi}{6}-\thetat\right)(e^{-iq_1}-e^{iq_2})\right)
\\ \notag \\
\Gamma_D^{12}&=0 \\
 \Gamma_D^{11}&=2\Psi_D \left( \cos^2\left(\thetat\right)(\cos q_2- \cos(q_1+q_2))+\cos^2\left(\frac{\pi}{3}-\thetat\right)(\cos q_1 -\cos q_2)+\cos^2\left(\frac{2\pi}{3}-\thetat\right)( \cos(q_1+q_2)-\cos q_1)\right)
\\ \notag \\
 \Gamma_E^{12}&= -\Psi_E \left(\cos^2\left(\frac{\pi}{2}-\thetat\right)e^{iq_2}+\cos^2\left(\frac{\pi}{6}-\thetat\right)e^{-i(2q_1+q_2)}+ \cos^2\left(\frac{5\pi}{6}-\thetat\right)e^{-iq_2}\right)\\
\Gamma_E^{11}&=\frac{3}{2}\Psi_E 
\\ \notag \\
\Gamma_F^{12}&= \Psi_F \left( \cos^2\left(\frac{\pi}{2}-\thetat\right)(e^{-i(2q_1+q_2)}-e^{-iq_2})+\cos^2\left(\frac{\pi}{6}-\thetat\right)(e^{-iq_2}-e^{iq_2})+\cos^2\left(\frac{5\pi}{6}-\thetat\right)(e^{iq_2}-e^{-i(2q_1+q_2)})\right)\\
\Gamma_F^{11}&= \Psi_F \left( \cos^2\left(\frac{\pi}{2}-\thetat\right)(e^{iq_1}-e^{-iq_1})+\cos^2\left(\frac{\pi}{6}-\thetat\right)(e^{iq_2}-e^{-iq_2})+\cos^2\left(\frac{5\pi}{6}-\thetat\right)(e^{-i(q_1+q_2)}-e^{i(q_1+q_2)})\right)
\\ \notag \\
\Gamma_G^{12}&= -\Psi_G \left(\cos^2\left(\frac{\pi}{2}-\varphi-\thetat\right)e^{-i(2q_1+2q_2)}+ \cos^2\left(\frac{\pi}{2}+\varphi-\thetat\right)e^{-i(q_1+2q_2)}+\cos^2\left(\frac{\pi}{6}-\varphi-\thetat\right)e^{iq_1}\right)\notag\\
& \;\;\;\; -\Psi_G \left(\cos^2\left(\frac{\pi}{6}+\varphi-\thetat\right)e^{i(q_1+q_2)}+ \cos^2\left(\frac{5\pi}{6}-\varphi-\thetat\right)e^{i(q_2-q_1)}+\cos^2\left(\frac{5\pi}{6}+\varphi-\thetat\right)e^{-2iq_1}\right)\\
\Gamma_G^{11}&=3\Psi_G 
\end{align}
where $\varphi=\tan^{-1}\frac{\sqrt{3}}{5}$ arises from the the orientation of the fourth nearest-neighbor terms $\Gamma_G$. After simplification, $\Gamma=\tau_0 \mathit{I}+\VEC \tau\cdot \sigma$, where 
\begin{eqnarray}
\tau_1&=&\cos^2\left(\frac{\pi}{2}-\thetat\right)\left(\Psi_C(\cos q_1-1)-\Psi_A\cos(q_1+q_2)-\Psi_E\cos q_2+\Psi_F(\cos(2q_1+q_2)-\cos q_2)\right)\notag\\
&&+\cos^2\left(\frac{\pi}{6}-\thetat\right)\left(\Psi_C(\cos(q_1+q_2)-\cos q_1)-\Psi_A-\Psi_E\cos (2q_1+q_2)\right)\notag\\
&&+\cos^2\left(\frac{5\pi}{6}-\thetat\right)\left(\Psi_C(1-\cos (q_1+q_2))-\Psi_A\cos q_1-\Psi_E\cos q_2+\Psi_F(\cos q_2-\cos(2q_1+q_2))\right)\notag\\
&&-\Psi_G \left(\cos^2\left(\frac{\pi}{2}-\varphi-\thetat\right)\cos(2q_1+2q_2)+ \cos^2\left(\frac{\pi}{2}+\varphi-\thetat\right)\cos(q_1+2q_2)+\cos^2\left(\frac{\pi}{6}-\varphi-\thetat\right)\cos q_1\right)\notag\\
&& -\Psi_G \left(\cos^2\left(\frac{\pi}{6}+\varphi-\thetat\right)\cos(q_1+q_2)+ \cos^2\left(\frac{5\pi}{6}-\varphi-\thetat\right)\cos(q_2-q_1)+\cos^2\left(\frac{5\pi}{6}+\varphi-\thetat\right)\cos 2q_1\right)\notag\\
\end{eqnarray}
\begin{eqnarray}
\tau_2&=&\cos^2\left(\frac{\pi}{2}-\thetat\right)\left(\Psi_C\sin q_1-\Psi_A\sin(q_1+q_2)+\Psi_E\sin q_2+\Psi_F(\sin q_2-\sin(2q_1+q_2))\right)\notag\\
&&+\cos^2\left(\frac{\pi}{6}-\thetat\right)\left(\Psi_C(\sin(q_1+q_2)-\sin q_1)-\Psi_E\sin(2q_1+q_2)+2\Psi_F\sin q_2\right)\notag\\
&&+\cos^2\left(\frac{5\pi}{6}-\thetat\right)\left(-\Psi_C\sin (q_1+q_2)-\Psi_A\sin q_1-\Psi_E\sin q_2-\Psi_F(\sin q_2-\sin(2q_1+q_2))\right)\notag\\
&&+\Psi_G \left(-\cos^2\left(\frac{\pi}{2}-\varphi-\thetat\right)\sin(2q_1+2q_2)- \cos^2\left(\frac{\pi}{2}+\varphi-\thetat\right)\sin(q_1+2q_2)+\cos^2\left(\frac{\pi}{6}-\varphi-\thetat\right)\sin q_1\right)\notag\\
&& +\Psi_G \left(\cos^2\left(\frac{\pi}{6}+\varphi-\thetat\right)\sin(q_1+q_2)+ \cos^2\left(\frac{5\pi}{6}-\varphi-\thetat\right)\sin(q_2-q_1)-\cos^2\left(\frac{5\pi}{6}+\varphi-\thetat\right)\sin 2q_1\right)\notag\\
\end{eqnarray}
\begin{eqnarray}
\tau_0 &=& \frac{3}{2}\left(\Psi_A+\Psi_B+\Psi_E+2\Psi_G\right)\notag\\
&& +\cos q_1 \left(-\Psi_B\cos^2\left(\thetat\right)+\frac{\sqrt{3}}{2}(2\Psi_D-\Psi_C)\sin 2\thetat\right)\notag\\
&& +\cos q_2 \left(-\Psi_B\cos^2\left(\frac{2\pi}{3}-\thetat\right)+\frac{\sqrt{3}}{2}(2\Psi_D-\Psi_C)\sin 2(\thetat+\pi/3)\right)\notag\\
&& +\cos (q_1+q_2) \left(-\Psi_B\cos^2\left(\frac{\pi}{3}-\thetat\right)+\frac{\sqrt{3}}{2}(2\Psi_D-\Psi_C)\sin 2(\thetat+2\pi/3)\right)
\end{eqnarray}

The \Gru{} parameters due to an isotropic biaxial strain in the $xy$ plane  can be obtained by 
summing up the uniaxial \Gru{} parameters with 
$\VEC{f}= (\cos \thetat, \sin \thetat, 0)$ and $\VEC{f} = ( \cos (\thetat + \pi/2), \sin(\thetat + \pi/2),0)$. Due to the identity $\cos^2 q+\sin^2 q=1$ for any $q$, we replace all occurrences of 
cosine squared terms in $\tau_1,\tau_2$ and $\tau_0$ by $1$ 
so that
\begin{eqnarray}
\tau_1|_{\rm iso}&=&-\Psi_A(1+\cos(q_1+q_2)+\cos q_1)-\Psi_E(2\cos q_2+\cos (2q_1+q_2))\notag\\
&&-\Psi_G \left(\cos(2q_1+2q_2)+\cos(q_1+2q_2)+\cos q_1+\cos(q_1+q_2)+\cos(q_2-q_1)+\cos 2q_1\right)\notag\\
\label{isotau1}
\end{eqnarray}
\begin{eqnarray}
\tau_2|_{\rm iso}&=&-\Psi_A(\sin(q_1+q_2)+\sin q_1)-\Psi_E\sin(2q_1+q_2)+2\Psi_F\sin q_2\notag\\
&&+\Psi_G \left(-\sin(2q_1+2q_2)- \sin(q_1+2q_2)+\sin q_1+\sin(q_1+q_2)+\sin(q_2-q_1)-\sin 2q_1\right)\notag\\
\label{isotau2}
\end{eqnarray}
\begin{eqnarray}
\tau_0|_{\rm iso} &=& \frac{3}{2}\left(\Psi_A+\Psi_B+\Psi_E+2\Psi_G\right)-\Psi_B(\cos q_1+\cos q_2+ \cos(q_1+q_2))
\label{isotau0}
\end{eqnarray}
It is evident that with an isotropic biaxial strain, only certain linear combinations of the CFCs can be isolated. 

\end{widetext}

\end{document}